\documentclass[letterpaper,twocolumn,10pt]{sig-alternate-10pt}
\usepackage{usenix,endnotes}
\usepackage{amsfonts}
\usepackage{amsmath}
\usepackage[usenames,dvipsnames]{xcolor}
\usepackage{enumitem}
\usepackage{setspace}

\usepackage[left=1in,top=1in,right=1in,bottom=1in,columnsep=0.25in]{geometry}
\usepackage[pdfstartview=FitH, bookmarksnumbered=true, bookmarksopen=true, colorlinks=true, citecolor=blue, colorlinks=blue, linkcolor= blue, urlcolor=blue]{hyperref}
\usepackage{epsfig,endnotes,graphicx,mwe,float,soul,url,xspace,verbatim}
\usepackage[font=small,labelfont=bf]{caption}
\usepackage[hang,scriptsize,tight,nooneline]{subfigure}

\usepackage{outlines}   

\newcommand{\cut}[1]{}

\newcommand{\cutatlastminute}[1]{}
\newcommand{\parab}[1]{\vspace{0.04in}\noindent{\bf #1} }
\newcommand{\eg}{{\it e.g.,}\xspace}
\newcommand{\ie}{{\it i.e.,}\xspace}
\newcommand{\sysnameNoSpace}{\textsc{Fetch}}
\newcommand{\sysname}{\sysnameNoSpace\xspace}

\newcommand{\loczurich}{Z\"{u}rich\xspace}
\newcommand{\loclahore}{Lahore\xspace}

\definecolor{lightgray}{gray}{0.95}
\newcommand\greybox[1]{%
        \vspace{4pt}%
        \par{\centering\colorbox{lightgray}{%
                \begin{minipage}{\columnwidth}#1\end{minipage}%
                }%
                \vskip 2pt%
                \vspace{4pt}%
}
}

\begin{document}
\setlength{\abovedisplayskip}{4pt}
\setlength{\belowdisplayskip}{4pt}
\setlength{\abovedisplayshortskip}{2pt}
\setlength{\belowdisplayshortskip}{2pt}
\setlength{\belowcaptionskip}{-16pt}
\baselineskip=12bp

\title{\vspace{-.5in}\LARGE \bf Measuring and exploiting the cloud consolidation of the Web}

\author{
{\rm Debopam Bhattacherjee}\\
ETH Z\"{u}rich
\and
{\rm Muhammad Tirmazi}\\
Harvard University
\and
{\rm Ankit Singla}\\
ETH Z\"{u}rich
} 

\date{}

\maketitle
\section*{Abstract}
{\em We present measurements showing that the top one million most popular Web domains are reachable within $13$~ms (in the median) from a collective of just $12$ cloud data centers. We explore the consequences of this Web ``consolidation'', focusing on its potential for speeding up the evolution of the Web. That most popular services reside in or near a small number of data centers implies that new application and transport technologies can be rapidly deployed for these Web services, \emph{without} the involvement of their operators. We show how this may be achieved by orchestrating a handful of reverse proxies deployed in the same data centers, with new technologies deployed at these proxies being nearly as effective as deploying them directly to the Web servers. We present early measurements of this approach, demonstrating a $>$$50\%$ reduction in Web page load times for users with high latencies to Web servers. We also show that this model, using a small number of proxies, can be surprisingly competitive with extensive CDN deployments, especially in geographies with high last-mile latencies.
}




\section{Introduction}

%


Where does the Web live? While Internet hypergiants like Google and Facebook run their own extensive infrastructure to back their services, and others at the opposite end of the spectrum operate out of their own servers and clusters, a large number of today's popular Web services are supported by public cloud infrastructure. 

Prior studies have shed light on the increasing use of cloud services like Amazon Web Services (AWS) and Microsoft Azure by comparing IP addresses seen in traffic towards popular Web services to address blocks published by these cloud providers~\cite{he2013next,whoWasAkella, labovitzAmazon}. We take the opposite perspective: given the well-established usage of cloud platforms by popular Web services, we conduct latency measurements to Web services \emph{from} these platforms, thus effectively conducting a census of Web services in terms of their proximity to AWS and Azure.

One key finding from our measurements is that the top million most popular Web domains (per Alexa~\cite{alexa}) are reachable in under $13$ milliseconds (in the median) from at least one vantage point out of just $12$ we deployed in Amazon's EC2 and Microsoft's Azure. While the replication of Web services via content distribution networks certainly decreases latency to them, including from EC2 and Azure, such replication alone does not entirely explain our results --- per a recent analysis, less than $10\%$ of the $32$,$000$ most popular Web sites use CDNs, with CDN usage declining rapidly with decrease in popularity~\cite{gilad2016cdn}. Thus, we conclude that a substantial fraction of the one million most popular Web services is hosted in or near a small number of cloud data centers. We refer to this observation as the cloud consolidation\footnote{We deliberately avoid the term ``centralization'', as it is already in use in a different, albeit related context --- coarsely, the idea that the attention of Web users is increasingly controlled by a small number of companies~\cite{webCentralization}.} of the Web.

The extent of cloud consolidation revealed by our measurements is much larger than is obvious from past work using address matching. For instance, He et al.~\cite{he2013next} found that only ``$4\%$ of the Alexa top million use EC2/Azure''. While we find that this percentage has also moved up in the intervening years --- from $4\%$ in $2013$ to $7\%$ in $2018$ ---  it still understates the cloud consolidation of the Web. Our latency-based measurements thus allow an examination beyond the tight constraint of matching IP addresses, revealing that a large fraction of Web services not hosted directly on the largest cloud platforms are still hosted (or at least replicated) somewhere in their near vicinity. 

We further observe that a few key Web hosting providers are the primary cause of this massive consolidation. 
Although the total number of providers hosting domains near public cloud is quite large, very few of them have their presence across multiple cloud data centers and consistently host a large fraction of these domains. Just $5$ key Web hosting providers host more than $85\%$ of such near-cloud domains across AWS locations.

We also explore an interesting implication of this consolidation: its utility for evolving the Web. While Web and Internet technology is advancing at a rapid pace with the goal of improving responsiveness and user experience, there is a long tail of still popular Web services which lags the technology adoption curve -- even GZip use is far from universal~\cite{agababov2015flywheel}, and newer technologies like WebP see adoption rates under $1\%$~\cite{agababov2015flywheel}. Likewise, while modern transport protocols like BBR~\cite{bbr} and QUIC \cite{quic} may see rapid uptake and interest at the likes of Google and Akamai, $14\%$ of Web servers still use a TCP initial congestion window of $2$ or $4$ segments~\cite{iwIMC17}, despite $10$ segments being the IETF recommendation since more than $5$ years~\cite{ietfTCPwin}. Further, as we shall see later, technology adoption is even slower in the developing World.

If we can place vantage points near \emph{most} Web servers, like our measurements show, we can use these as proxies for Web content. Using new Web and Internet technologies at such near-server proxies can often be nearly as effective as using them directly at the Web servers, and does not require \textit{any} involvement from the operators of Web services. While proxy-based solutions for speeding up the Web are already well-studied~\cite{agababov2015flywheel,parcelSivakumar,netravali2015mahimahi, sehati2016network, operaturbo, amazonsilk, freeBasics}, our measurements reveal a simple and cheap \emph{deployment model} for such proxies. At the small latencies between the Web servers and the proxies in our proposed deployment, old transport and Web stacks incur small overheads, and wide-area connectivity between these proxies and clients can benefit from modern network stacks. Thus, a small number of proxies can speed up the Web's evolution, and substantially improve performance for many clients and services today.

We build a prototype of this approach, \sysname, and evaluate it using both emulation across network configurations with varying latency, bandwidth, and loss; as well as with a small number of Internet vantage points, finding that it can improve Web performance by more than $50\%$ for users with high latencies to Web servers. We also show that \sysname can be surprisingly competitive against Web page delivery over an extensive, well-tuned content delivery network, achieving page load times within $5\%$ of the CDN-based approach.

\begin{figure*}
        \begin{center}
                \includegraphics[width=\textwidth]{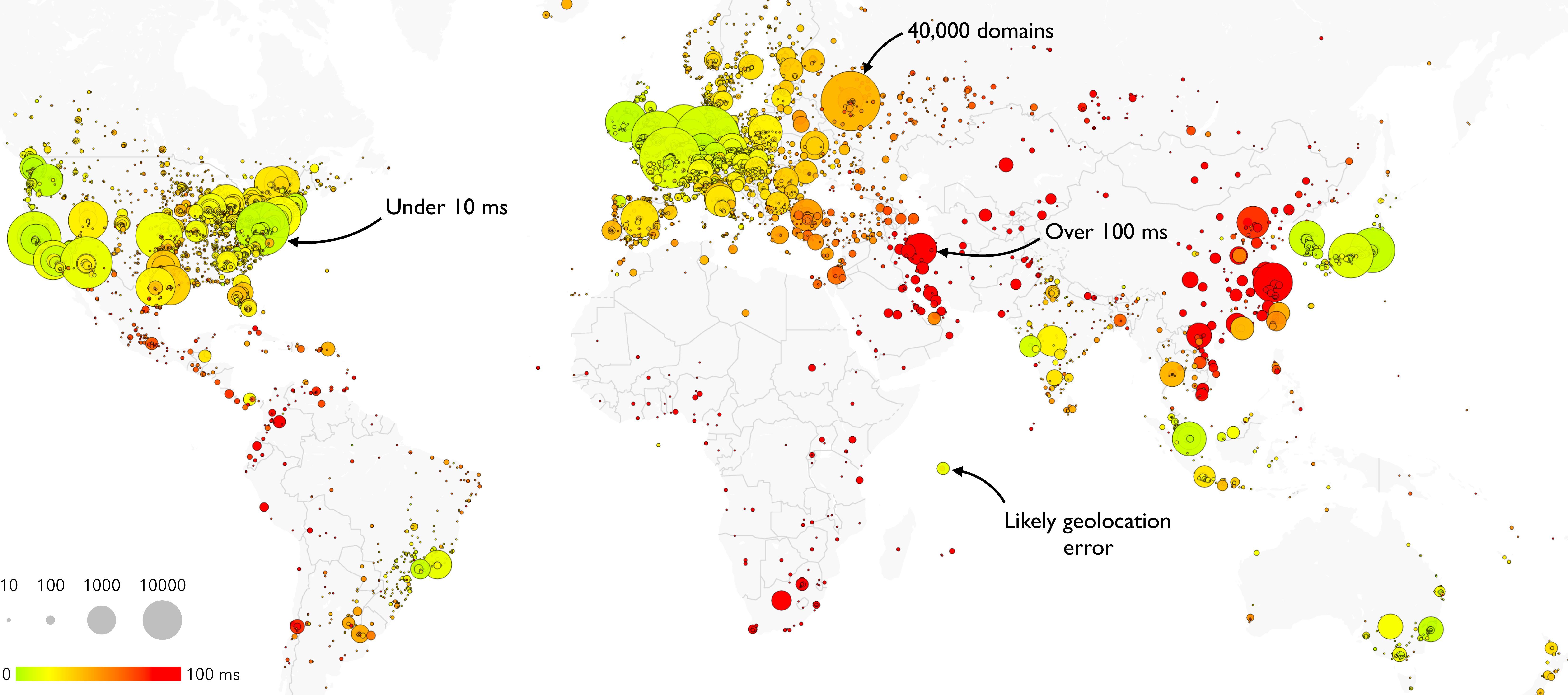}
                \caption{\em Latencies from EC2 to Web servers. As the legend indicates, circle sizes reflect the number of servers geolocated to that location, and circle color reflects average latency across those domains, green (lighter in grayscale) being lowest (under $5$~ms), and red (darker) being the highest ($100$$+$~ms). Note that for visual clarity, \textbf{circle sizes are proportional to the cube-root of the number of domains} geolocated to a location.}
                \label{fig:globalViz}
        \end{center}
\end{figure*}

\section{Cloud consolidation of Web services}
\label{sec:consolidation}




We quantify round-trip latency to Web services from $14$ Amazon EC2 data centers. From each of these $14$ data centers, we measured round trip times to Web servers hosting the top one million most popular Web sites (per Alexa's list~\cite{alexa}) in May $2018$. We used \texttt{hping}~\cite{hping} to conduct our RTT measurements, allowing us to send TCP SYN packets to the Web servers and record when the TCP SYN-ACKs were received at our Amazon nodes. We refer to the least of the $14$ measurements for each Web site as the ``latency from EC2''. Of course, for services which use geo-replicated deployments with anycast or DNS redirects, our $14$ measurements may not correspond to the same physical Web server, but this does not hinder our goal of quantifying the latencies to these Web \emph{services} from each of our measurement sites.

Fig.~\ref{fig:globalViz} shows a visualization of latencies from EC2 to the one million most popular Web domains. In addition to the \texttt{hping} measurements, this visualization also uses geolocation to map the Web servers which correspond to the smallest measured latency from EC2. For this purpose, we used MaxMind's free geolocation product~\cite{geolite}, rounding locations to one decimal degree. While there are bound to be some geolocation errors, it is unlikely that these change the visualization significantly, as it is broadly in line with our latency measurements (which we shall detail shortly). Each circle's size denotes the number of services in the top million that were geolocated to the circle's center. For clarity, circle sizes are scaled to the cube-root of the number of domains geolocated to a location, instead of linear. Color denotes average latency across domains geolocated to that location, with darker / warmer colors being worse. This visualization rounds all measurements larger than $100$~ms ($<$$3\%$ of the data) to $100$~ms for greater resolution for the rest of the data.

\begin{figure}
        \centering
        \includegraphics[width=3in]{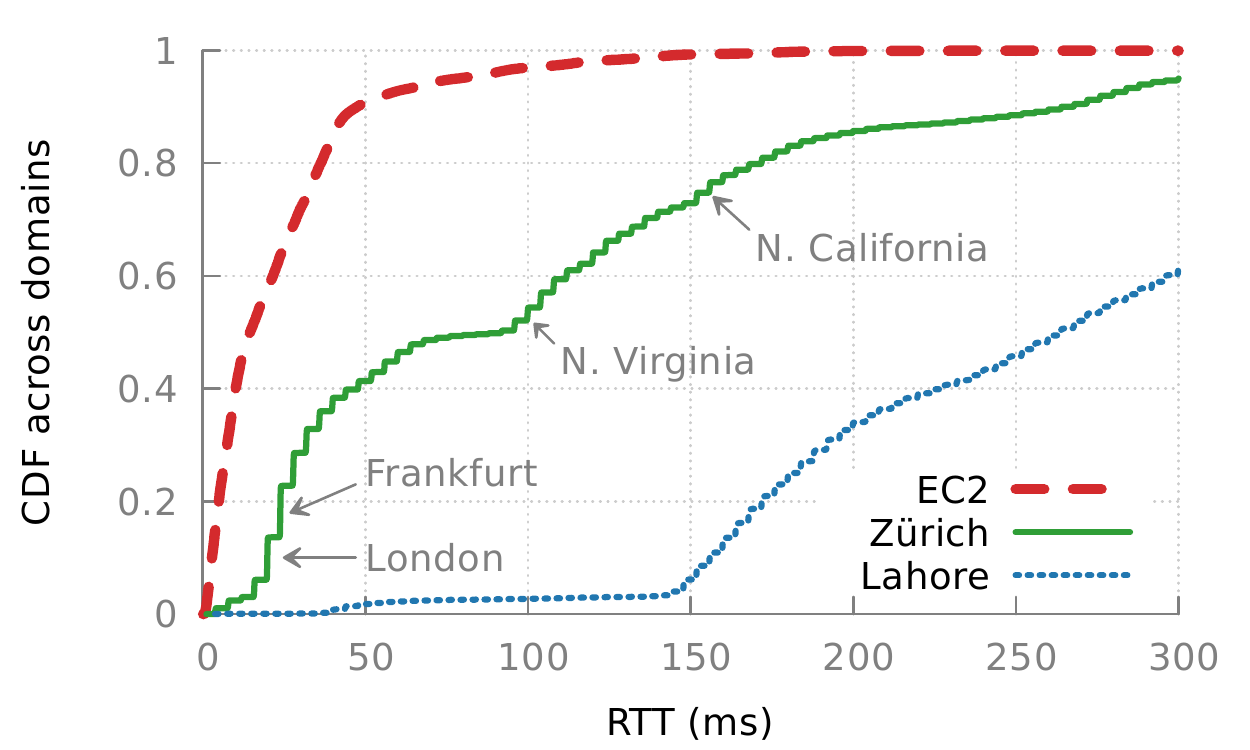}
        \caption{\em EC2 achieves lower latencies to services than \loclahore and \loczurich.}
        \label{fig:rttmeasure:RTTCDF}
\end{figure}

As is clear from the figure, the EC2 view of services is far from uniform across the globe, with most being consolidated in a small number of locations: a few large circles contain most of the mass\footnote{The cubic scaling understates this concentration.}; and most services are reachable within a small latency from our EC2 nodes: most of the mass in the figure is green-yellow. 

We also measured RTTs to the same Web services from two university-based nodes, one in Europe (\loczurich), and another in Asia (\loclahore). Measurements from \loclahore were much slower, and we used a random sample of $\sim$$50$,$000$ Web sites to test. As Fig.~\ref{fig:rttmeasure:RTTCDF} shows, latency is lowest from EC2 and highest from our Asia-based client, with the medians ($90$\textsuperscript{th}-percentiles) being $15$~ms ($48$~ms), $92$~ms ($264$~ms), and $264$~ms ($468$~ms) from EC2, \loczurich, and \loclahore respectively. Thus, median RTT from EC2 is $6$$\times$ smaller than from \loczurich, and $18$$\times$ smaller than from \loclahore. Note that both of these clients are university-hosted and ``real'' end-user connectivity may be worse. We omit visualizations equivalent to Fig.~\ref{fig:globalViz} for \loczurich and \loclahore, but note that the structure of the map is nearly identical in each case, and for \loczurich, the map is largely red (high latency) outside of Europe; while for \loclahore, it is almost entirely red. Several interesting features are evident in Fig.~\ref{fig:rttmeasure:RTTCDF}:

\begin{itemize}[leftmargin=11pt]
\vspace{-5pt}
\setlength\itemsep{0em}

        \item Measurements from \loczurich reveal a step structure, with substantial parts of the distribution contained in some steps. These steps correspond to Web servers consolidated in data centers. We identify and point out several such ``steps'' in Fig.~\ref{fig:rttmeasure:RTTCDF}: Frankfurt, London, N. Virginia, and N. California\footnote{Amazon does not provide more specific location identifiers.}. The Frankfurt and London steps respectively contain $9\%$ and $7.5\%$ of the entire distribution's mass within a half second interval each. 
        \item The ``plateau'' in latencies from \loczurich ($60$-$100$~ms), where there are few measurements is mostly due to the trans-Atlantic latency from \loczurich to servers in the Americas.
        \item From \loclahore, most Web servers are far, with $94\%$ being more than $150$~ms away. We conjecture that the step-characteristic of the \loczurich measurements is absent here due to greater latency variations across longer paths.
\end{itemize}

\subsection{Are these observations stable?}
\label{subsec:stableMapping}


We also assess whether these measurements are stable in two senses: (a) Is the large latency to some servers (the tail in Fig.~\ref{fig:rttmeasure:RTTCDF}) simply an artifact of variance which could be removed by taking the minimum across multiple measurements to each server? and (b) Is the measured latency between an EC2 location and a Web server consistent across longer periods of days?

\parab{(a) Variance:} It is reasonable to expect that out of several million \texttt{hping}s, some result in large latencies simply because of some servers being temporarily overloaded or due to transient network congestion. Thus we also conducted repeat measurements to a sample of servers. For each $5$~ms bucket on the $x$-axis in Fig.~\ref{fig:rttmeasure:RTTCDF}, we randomly sampled $100$ servers for which one-shot RTTs were in that bucket, and collected $10$ measurements to each over the course of several hours. For each server, we consider the minimum of these measurements as its definitive RTT. These definitive RTTs are indeed lower than one-shot measurements, but the differences are on the order of $10\%$ across the entire range of RTTs seen in Fig.~\ref{fig:rttmeasure:RTTCDF}. Thus, the tail in Fig.~\ref{fig:rttmeasure:RTTCDF} is not just an artifact of variance.

\parab{(b) Long-term consistency over days and weeks:} We created a mapping of each Web server to its closest EC2 node. After a period of one week, we repeated measurements between Web servers and their mapped EC2 nodes, and quantified the (absolute) changes in RTT across server-EC2-node-pairs between these measurements. 
The changes are typically small: $2$~ms in the median, and $22$~ms at the $80\%$\textsuperscript{th} percentile. Further, we find that for Web servers to which the initial RTT measurement is small, this variation is even smaller -- across Web servers for which our first RTT measurement was under $20$~ms, the $80$\textsuperscript{th} percentile change in RTT is only $4.3$~ms. These results are not surprising: we expect that the way a service is hosted does not typically change on a weekly basis.



\begin{figure*}[t]
  \centering
  \subfigure[]{\label{fig:azure-ec2}
    \includegraphics[width=3in]{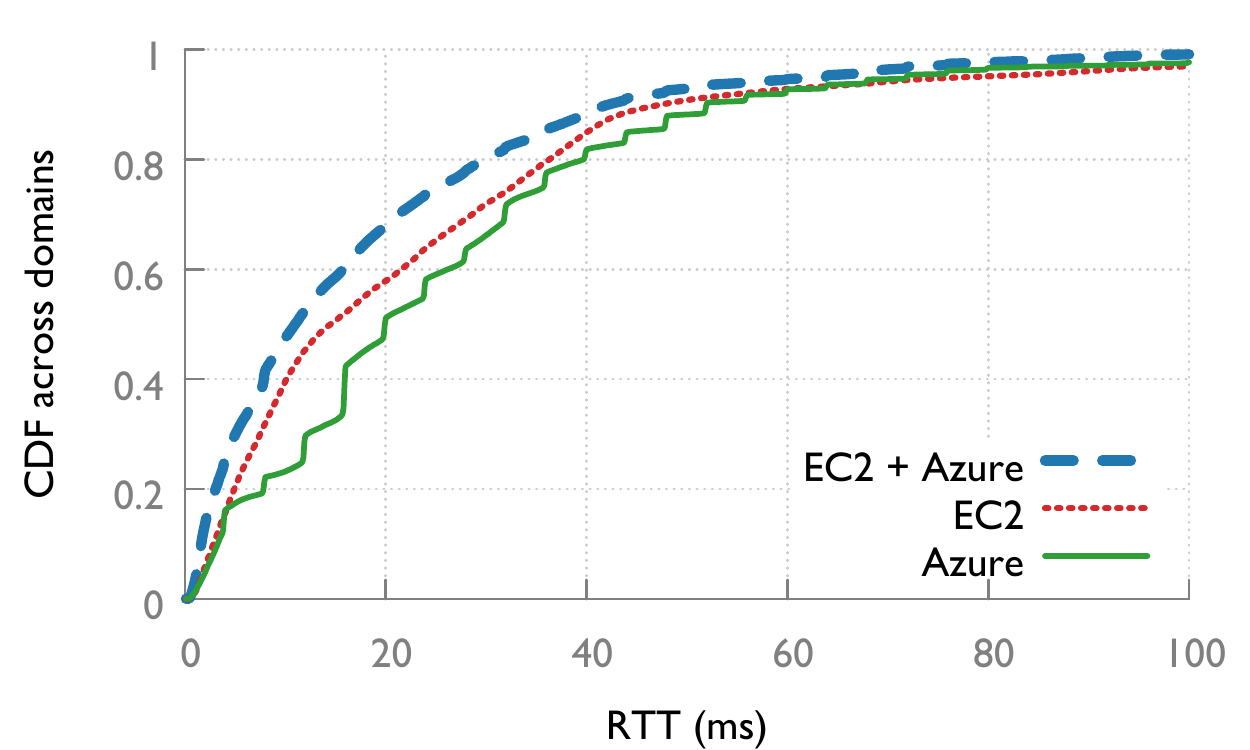}}
  \subfigure[]{\label{fig:addingLocations}
    \includegraphics[width=3in]{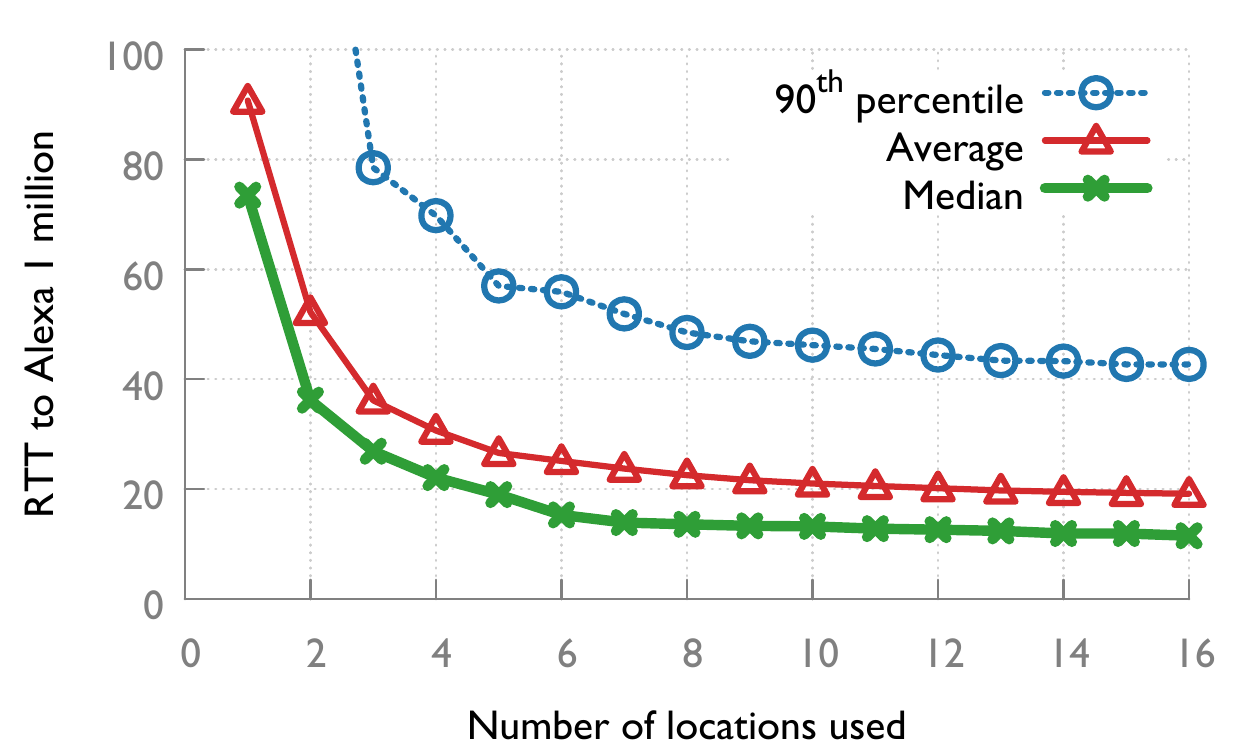}}
  \subfigure[]{\label{fig:selectelFilters}
    \includegraphics[width=3in]{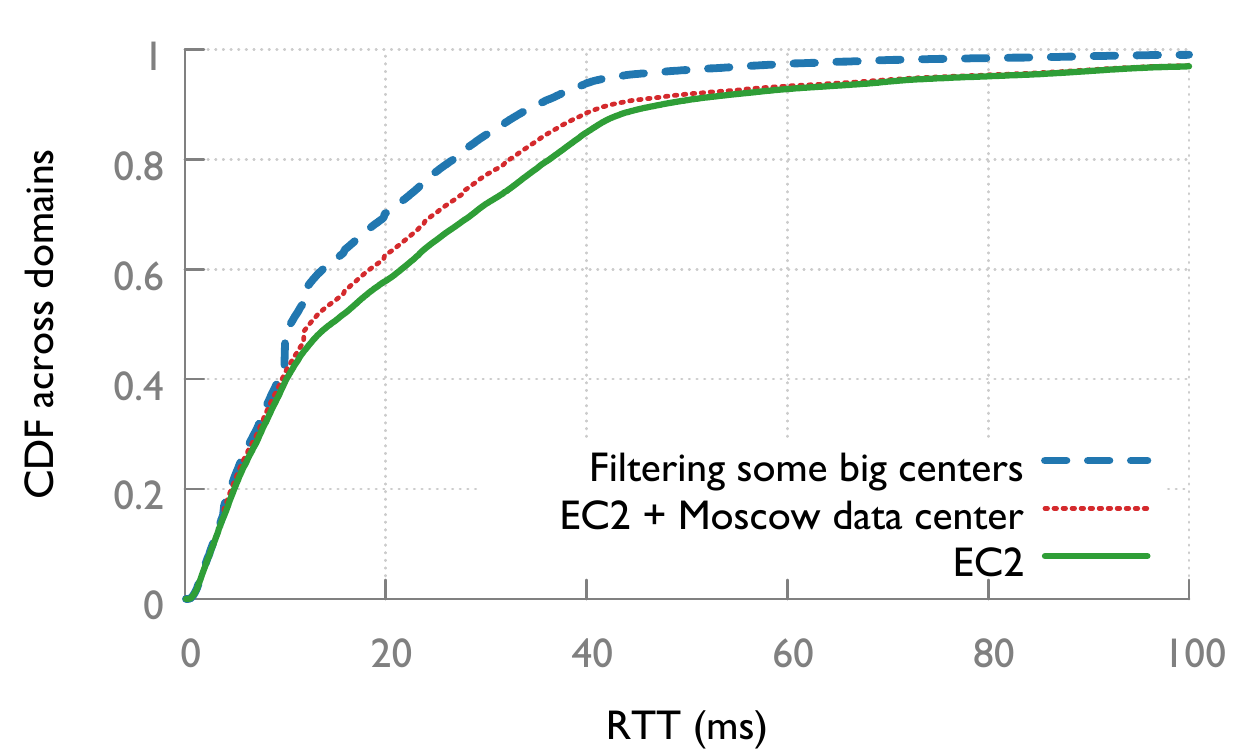}}
  \subfigure[]{\label{fig:longterm}
    \includegraphics[width=3in]{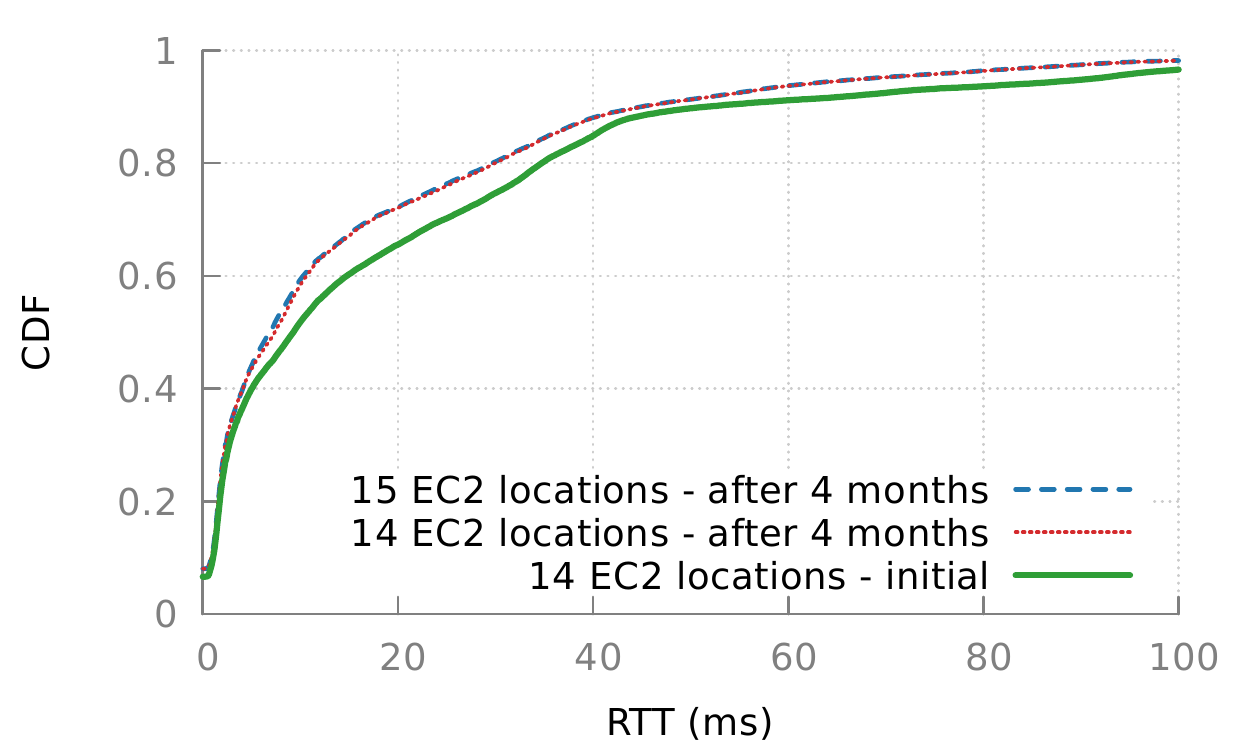}}
   \vspace{-8pt}
  \caption{\em Lowering latencies further: (a) latencies from EC2, Azure, and both together; (b) beyond a small number of chosen locations, adding more EC2 and Azure locations yields small marginal gains; and (c) adding measurements from a Moscow data center reduces latencies further, as does filtering out Chinese domains, and Web servers geolocated to $5$ additional locations; and (d) latencies from EC2 have improved in the last $4$ months for the top $100$,$000$ domains.}
  \label{fig:lowerFurther}
  \vspace{6pt}
\end{figure*}

\subsection{Lowering latency even further}
\label{subsec:lowerFurther}


While latencies from EC2 to Web servers are small, we also investigate ways of lowering them even further.

First, we add similar measurements from Microsoft's Azure platform to compare with those from EC2. On Azure, $12$ data center locations were accessible to us. As Fig.~\ref{fig:azure-ec2} shows, latencies from Azure are larger (by $38\%$ in the median), but using both together (such that the reported latency is the least across all $26$ measurements for each Web site) can reduce latencies a bit further to $11$~milliseconds in the median (\ie a $24\%$ reduction).


We also attempt to quantify the contribution of additional locations to these reductions in latency. Given a candidate set of $N_l$ locations, and a ``budget'' of $l$ locations we can choose, we would like to pick the $l$ locations such that they minimize a latency objective, such as the median, average, or $95$\textsuperscript{th} percentile. A trivial reduction from the facility location problem~\cite{facilityLocation} to this problem establishes its NP-Hardness. While our brute-force attempts failed even with this small number of candidate sites ($N_l = 26$), we used a simple iterative heuristic: choose a subset of $20$ locations (a size for which brute-force suffices) and keep the best set of $15$ locations, discarding the rest. Repeating this procedure a few times reveals that the incremental benefit of adding locations beyond $15$ is minimal. This characteristic is shown in Fig.~\ref{fig:addingLocations}. Using the $12$ locations that minimize the average, we find that the median, average, and $90$\textsuperscript{th}-percentile latencies are $13$, $20$, and $44$~milliseconds respectively.

However, we observe that the main reason additional locations do not help is their natural redundancy -- most Azure locations are near some EC2 location, while neither covers some parts of the globe. As Fig.~\ref{fig:globalViz} shows, the Web servers for which we see high latencies are also consolidated in or near a small number of big population centers, including Beijing, Istanbul, Johannesburg, Moscow, Shanghai, and Tehran. Each of these also corresponds to data center locations, just not ones available to us with EC2 or Azure. For instance, there are Telehouse data centers~\cite{telehouse} in Johannesburg and Istanbul. Amazon and Azure do in fact offer locations in China, but using these requires registering a Chinese legal entity~\cite{amazonChinaSignup}. 

We assess the impact of adding presence at these locations in three ways: (a) adding measurements from a Moscow data center; (b) filtering out Chinese prefixes; and (c) coarsely emulating the inclusion of $5$ additional locations by geolocating high-latency servers and setting measurements corresponding to the $5$ largest locations (\ie with most domains geolocated to those locations) to $10$~milliseconds, in line with our observation that latencies within the vicinity of a data center are usually under $10$~ms. Note that this emulation is likely to understate the impact of adding more locations, as we cannot account for reduction in latencies for other Web sites not in the immediate vicinity of these emulated locations.


Fig.~\ref{fig:selectelFilters} shows latencies from EC2, with the addition of measurements from the Moscow data center, and further, with the filtering of Chinese domains and $5$ other (city-level) locations. The filtered latencies are $27\%$ lower in the median than those from EC2, and $45\%$ lower in the $95$\textsuperscript{th}-percentile. Thus, adding a small number of locations in a targeted manner may offer substantial further reductions on the already small latencies we measure.

\begin{figure}
\begin{center}
\includegraphics[width=3in]{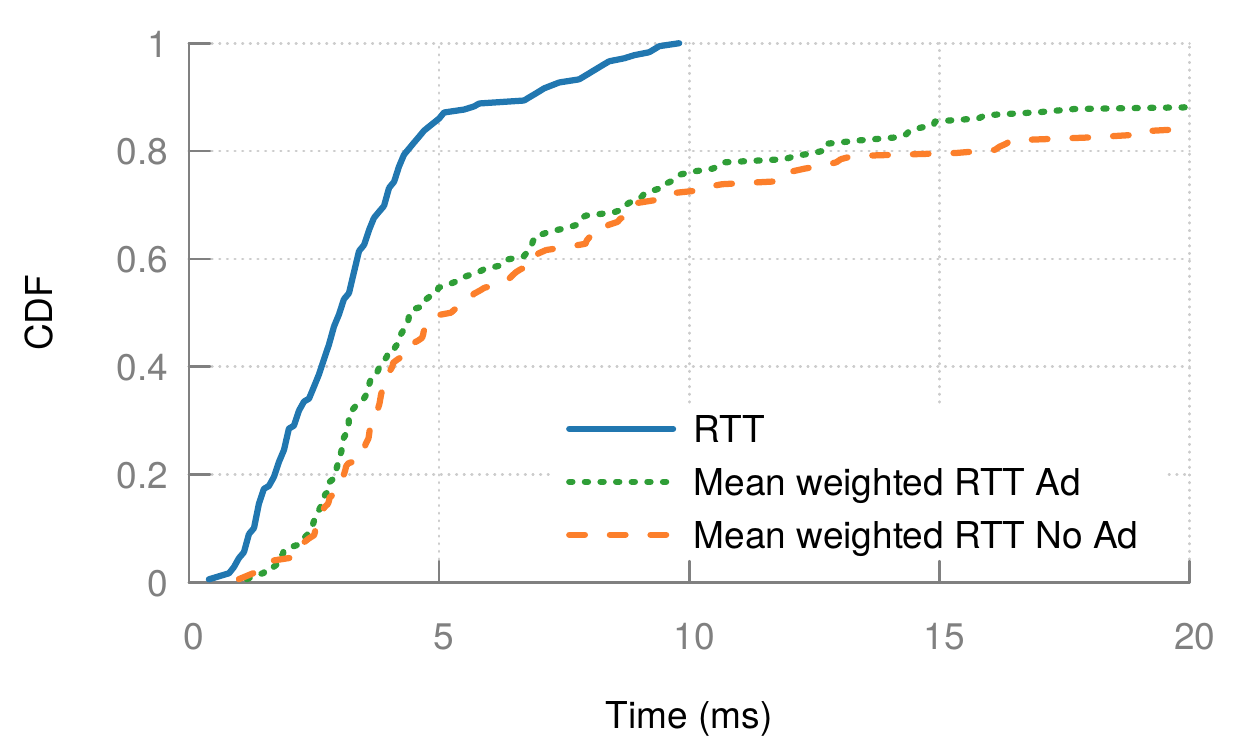}
\caption{\em Mean weighted RTTs to non-origin domain servers.\vspace{-4pt}}
\label{fig:results:wrtt}
\end{center}
\end{figure}

\subsection{Further consolidation over time}
\label{subsec:temporalconsolidation}
We measured the latencies from EC2 to the top $100$,$000$ domains out of the previously measured $1$ million domains after a period of $4$ months, in September $2018$, in order to analyze the consolidation happening over time. We limited this analysis to the top $100$,$000$ domains as we observed that less popular domains lower in the list have a high  long-term churn rate 
which would result in measurements to very different sets of services. 
Our later measurements also cover an additional, relatively new EC2 data center in Paris. Fig.~\ref{fig:longterm} shows that there has been a futher significant consolidation during this period even if we account for only the $14$ data centers that we considered the previous time. The median ($95$\textsuperscript{th}-percentile) latency improved by $20\%$ ($25\%$) falling from $9.3$~ms ($91.2$ ms) to $7.4$~ms ($68.2$~ms). If we take the Paris data center into account, latency has dropped by $26\%$ in the median as well as the $95$\textsuperscript{th}-percentile.

\subsection{What about non-origin domains?}
\label{subsec:thirdpartydomains}

Most Web services today do not operate as monoliths, with the content of Web pages often composed of responses from many servers across many domains besides the origin. It is thus plausible that we measure small latencies to the origin, but the non-origin domains serve a large fraction of a Web page's content and are reachable only at high latencies. We thus measure latencies to these non-origin domains and compare them to those for the origin servers.

We restrict these measurements to popular sites for which we observe small latencies for the origin. We use the top $200$ Alexa Web pages (excluding Google pages for greater diversity) for which RTTs from Azure's US WEST 2 data center are less than $10$ms. For each Web page, we record the number of bytes fetched from each non-origin domain and the RTT to that domain. We then compute for each page, mwRTT, the mean RTT weighted by the number of bytes across these domains. 

Fig.~\ref{fig:results:wrtt} shows RTTs to the origin servers and mwRTTs including and excluding ad servers (identified using Easylist~\cite{easylist}, which is used by popular ad blockers). Measurements with and without ad servers filtered out are broadly similar. The median mwRTT to servers including ad servers is $4.4$~ms, a bit higher than the RTT to the origin server, but still small in the absolute.


\greybox{\parab{Summary:} Our measurements and analysis reveal massive consolidation of Web services in or near a few big cloud data centers, with most services being reachable at low latencies from at least one of a handful of vantage points in these data centers.}


\begin{figure*}[t]
  \centering
  \subfigure[]{\label{fig:dcwise_providers}
    \includegraphics[width=3in]{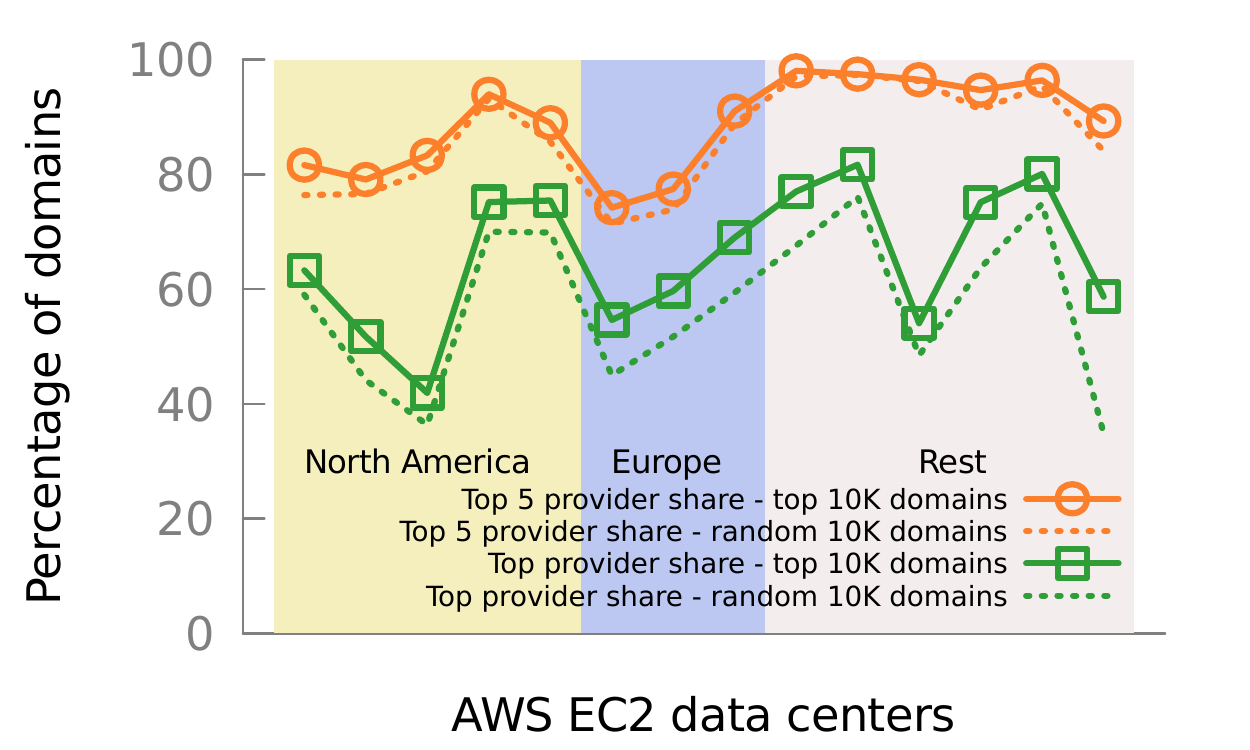}}
  \subfigure[]{\label{fig:3dc_top5share}
    \includegraphics[width=3in]{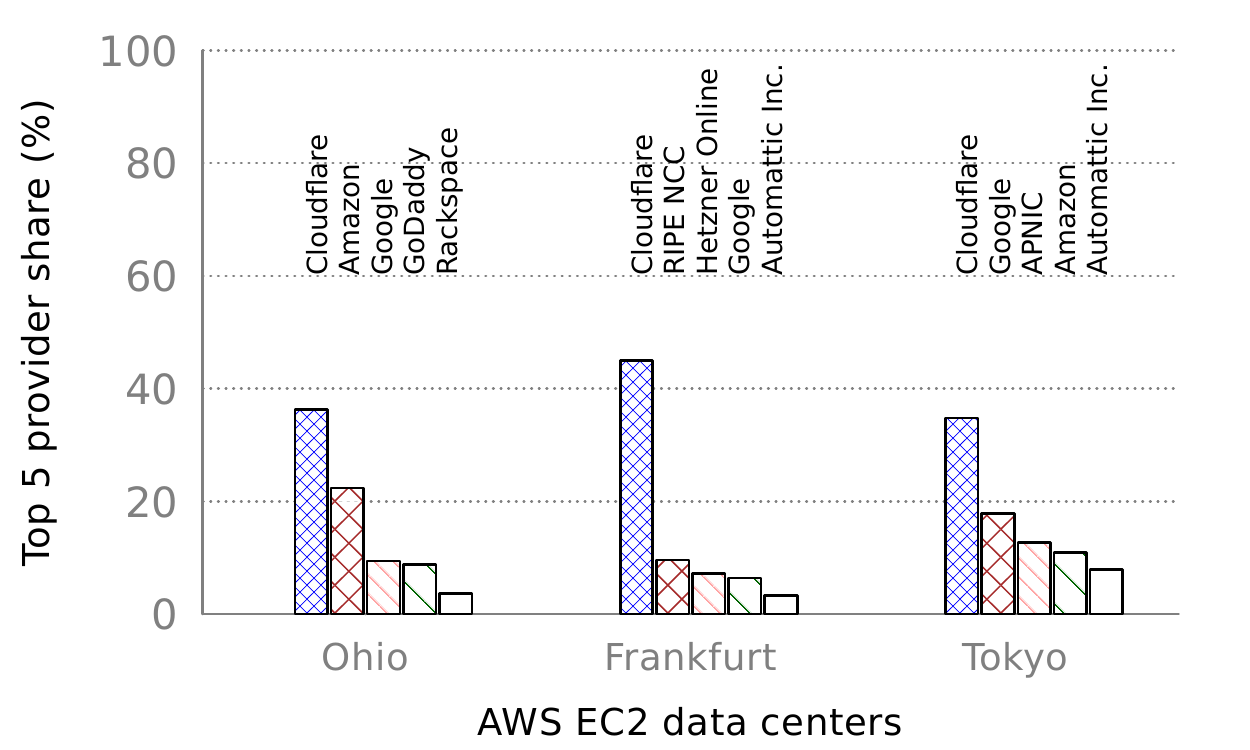}}
  \setlength{\belowcaptionskip}{-15pt}
  \caption{\em Few large Web hosting providers exist: across EC$2$ data centers (a) top $5$ ($1$) hosting providers are responsible for at least $71\%$ ($34\%$) of the domains ; and (b) top $5$ providers' relative dominance varies (Frankfurt vs. Tokyo) while the aggregate share is still significant.
  }
  \label{fig:providers}
  \vspace{6pt}
\end{figure*}

\section{What is behind cloud consolidation?}
\label{sec:providers}

We investigated how the primary domains, which are consolidated in or near the cloud data centers, are hosted. For each of the $14$ EC$2$ locations that we consider in \S\ref{sec:consolidation}, we examined the top $10$,$000$ and random $10$,$000$ domains which have less than $15$~ms RTT from the data center. For each EC2 site, we used \texttt{whois}~\cite{whois} to query the organization names (orgnames) hosting the responding servers. Certain organizations have varying orgnames across domains and locations. One such example is Amazon~\cite{amazon} using orgname \texttt{Amazon.com\-Inc.} as well as \texttt{Amazon\-Technologies\-Inc.}. We verified that Amazon Technologies, Inc. operates as a subsidiary of Amazon.com Inc.~\cite{amazon_subsidiary}. We mapped such related orgnames to unique organizations for our results.

\subsection{A few key providers per location}
\label{subsec:provider_each_dc}
In Fig.~\ref{fig:dcwise_providers} we plot the percentage of domains with valid orgnames hosted by the top $5$ (also, only the top) Web hosting providers across the $14$ EC2 data centers both for the top and random $10$,$000$ domains. As is evident, the share of the most significant provider, which is consistently Cloudflare~\cite{cloudflare} across our measurements, varies between $34\%$ (Tokyo) and $76\%$ (Sao Paulo) for random domains close to the cloud. For top domains as well, Cloudflare is the most significant provider consistently, with domain share varying between $42\%$ (Ohio) and $82\%$ (Sao Paulo). If we consider the top $5$ providers for each location, the range varies between $71\%$ (Frankfurt) and $97\%$ (Mumbai) for random domains, and $74\%$ (Frankfurt) and $98\%$ (Mumbai) for top domains. Nowhere is the share of top $5$ providers less than $71\%$. This shows that the consolidation of the Web near public cloud is a result of few key Web providers hosting a large chunk of these domains close to the cloud. 

It is also evident from Fig.~\ref{fig:dcwise_providers} that the aggregate shares of the top $5$ providers are higher, in general, for locations outside North America or Europe. We further observe that the average number of providers who host at least $1\%$ of the domains tested for per location, is $8.4$ for North America and Europe while it falls below $6$ for the rest of the world. This implies that the shares of the top $5$ providers are less across North America and Europe as there are a handful of other hosting providers present in those locations with significant deployments in or close to the public cloud.


\subsection{Provider share across EC2 sites}
\label{subsec:provider_across_dc}
For each of the $3$ regions (North America, Europe and rest), we examine the EC2 location where the share (for random domains) of the top provider (CloudFlare) is the lowest. Fig.~\ref{fig:3dc_top5share} shows the domain share of each of the top $5$ providers across these locations. In case of Frankfurt, Cloudflare has significantly higher share than the rest ($4.7\times$ the $2$\textsuperscript{nd} major provider); while, in case of Tokyo, the shares are somewhat more uniform ($2\times$ the $2$\textsuperscript{nd} major provider). However, as pointed out in the previous section, CloudFlare is consistently the dominant provider, and the aggregate share of the top $5$ exceeds $71\%$ of the domains tested. Note that RIPE NCC~\cite{ripencc} and APNIC~\cite{apnic} are governance entities and do not provide hosting services. The corresponding vertical bars in Fig.~\ref{fig:3dc_top5share} are the results of default \texttt{whois} setting; excluding them and including $6$\textsuperscript{th} top providers do not change the results significantly.

The top $5$ Web hosting providers responsible for more than $85\%$ of the top domains under consideration across EC2 locations are Cloudflare, Amazon, Google~\cite{google}, Akamai~\cite{akamai}, and Fastly~\cite{fastly} respectively. For the random domains, however, Google and Amazon change positions while the $4$\textsuperscript{th} and $5$\textsuperscript{th} positions are occupied by Automattic~\cite{automattic} (notable for WordPress.com~\cite{wordpress}) and GoDaddy~\cite{godaddy} respectively.
\greybox{
\parab{Summary:} Our measurements show that a few key Web hosting providers play a significant role behind the cloud consolidation of the Web that we observed. If these providers together capture an even larger percentage of the hosting market, we may observe further consolidation of the Web near the public cloud.
}

\begin{figure}
\begin{center}
\includegraphics[width=3in]{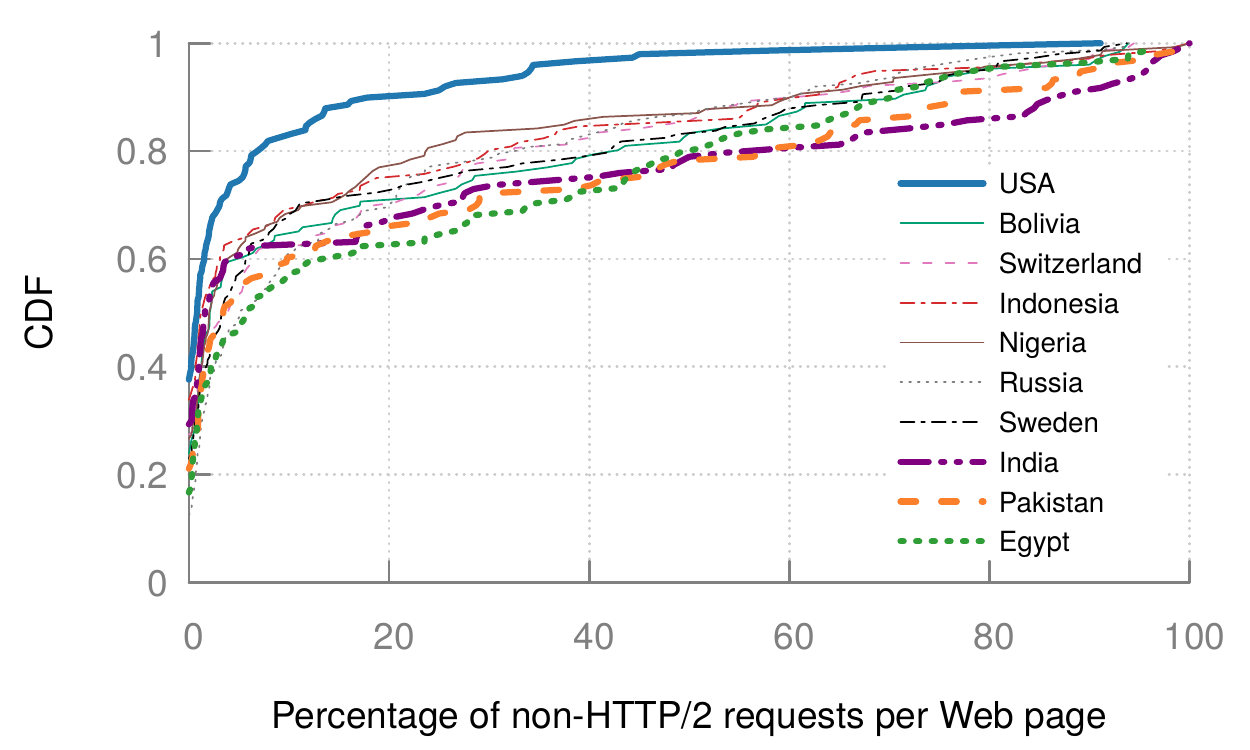}
\caption{\em For the top 200 Web pages, the percentage of non-HTTP/$2$ requests per Web page varies by more than $40\%$ across countries at the $75$-th percentile. HTTP/$2$ adoption is highest in the US.}
\label{fig:http2_adoption}
\end{center}
\end{figure}

\section{Speeding up Web evolution}
\label{sec:deploymentmodel}

Our measurement of Web consolidation in or near a few public cloud data centers opens up an interesting possibility: speeding up the evolution of Web services. We first motivate this use case (\S\ref{subsec:slowEvolution}), and then discuss how Web consolidation can help address it  (\S\ref{subsec:enableEvolution}).

\subsection{The slow-moving long tail}
\label{subsec:slowEvolution}

While the networking community is continually making significant performance improvements in the application and network stack, Web service operators beyond the industry leaders are often slow to adopt these technologies. Even universally supported technologies like using GZip for compression of Web page content are not ubiquitous~\cite{agababov2015flywheel}. Newer developments like WebP image compression~\cite{webp}, HTTP/$2$, and novel transport protocols, see even less penetration. 
Further, we find that the adoption of new technologies varies substantially across geographies. As one instance of this, we measure the adoption of HTTP/$2$ across different countries. For several countries, we evaluate the $200$ Web pages most popular in that country. Fig.~\ref{fig:http2_adoption} plots the CDF of the percentage of non-HTTP/$2$ (\ie HTTP/$1.1$ or lower) requests per Web page for various countries. For several countries, including Egypt, Pakistan, and India, more than $20\%$ of even this set of $200$ \emph{most popular} pages generate more than $50\%$ non-HTTP/$2$ requests. In comparison to the US, adoption is substantially lower elsewhere.

There is thus a large fraction of Web services which evolve slowly in terms of adopting new network and application layer advances, leading to sub-optimal performance and user experience. 

\subsection{Consolidation enables evolution}
\label{subsec:enableEvolution}

Inevitably, a large fraction of Web services will continue to operate with outdated and sub-optimal Web stacks far after the industry's leading edge has deployed more efficient technologies. However, cloud consolidation of the Web can help address this issue \emph{without} requiring the involvement of slow-adopter Web service operators.

Given that a small number of cloud vantage points can achieve proximity to most Web services, we can use these vantage points as proxies. Deploying new technologies at these proxies can often be nearly as effective as modifying the Web servers themselves, because at the small latencies between these proxies and Web servers, the inefficiencies from older stacks are minimal, and we can get the benefit of the new stacks' greater efficiency over (high-latency) wide-area connectivity between the clients and proxies. We outline a simple design, \textbf{\sysname}, along these lines, with two ingredients: (a) client software, which directs Web service requests to \sysname proxies which are nearest to those Web services; (b) \sysname proxies that receive and fulfill client requests.

\parab{Client-software:} A client needs to map any target Web service to the \sysname proxy closest to that service, and then send its request to this particular proxy. This mapping is a small file (few MB), which the client obtains from \sysname. Given the stability of the latencies between Web servers and their mapped proxies (\S\ref{subsec:stableMapping}), measurements, computations, transfers, and updates of these mappings are all infrequent (\eg once a day). In the occasional event that mappings are out of date (indicating that the Web service has migrated to a different data center, or its replica in the mapped location is slow or unavailable), there is a performance penalty, but the content can still be obtained -- the proxy simply fetches it like any other device on the Internet.

\parab{\sysname proxies:} Each \sysname proxy runs in a cloud data center. On receiving a page request, the \sysname proxy gets the content from a nearby Web server over (many) very short RTTs, potentially performs other optimizations (\eg compression), and delivers the content to the client in a batch, minimizing the number of long client-\sysname RTTs. This can be achieved using: (a) transport like QUIC~\cite{quic}, which eliminates the handshake in most cases\footnote{With \sysname, clients connect only to a handful of proxies for most requests, and thus rarely need a full handshake.}; (b) remembering and persisting initial window sizes between clients and proxies: the client logs the last observed TCP window size for incoming data, and sends it to the proxy in a cookie with every new request, thus letting the server start with this (usually) larger sending window; and (c) using the superior loss recovery of recent transport protocols like BBR~\cite{bbr}.

\sysname proxies also measure latency to popular Web services, and aggregate and process these to construct the proxy to Web-service mapping.

In a real deployment, \sysname proxies can be an adaptation of any of a number of existing proxy-based solutions. The proxies may themselves be replicated and use standard load balancing. For simplicity, we refer to one cloud deployment location as one proxy throughout.

\parab{Prototype implementation:} Our implementation uses a small piece of software at the client, which the browser uses as a local proxy through standard configuration mechanisms. It forwards Web requests to a suitable \sysname proxy based on the mapping discussed above, and as content is received from the \sysname proxy, it serves the content to the browser, which starts loading the page. Standard transport layer optimizations like larger window sizes~\cite{nanditaWindowSize} and handshake-free connections~\cite{TCPfastopen, quic} are used between the clients and the proxies. 
At each \sysname location, we maintain a pool of headless browsers (presently \texttt{PhantomJS}~\cite{phantomjsrelease}), to which incoming client requests are assigned. Depending on load, new instances may be launched on separate virtual machines, and requests may be spread across these via a load balancer. On receiving a page request, a browser instance fetches it as usual. Any content received is transmitted in parallel to the client, even as the browser processes it. 

Our goal is not to re-architect proxies, but to develop a simple way for deploying and orchestrating a small number of proxies at appropriate locations, in line with our measurements. We thus focus on a minimal implementation, relying on experience with deployed systems like Google's Flywheel~\cite{agababov2015flywheel} for software issues related to the substantial complexity of the Web ecosystem. 
\section{Evaluation: how much could this help?}
\label{sec:results}

We evaluate Web page performance with \sysname, using (known) transport optimizations, batching responses from the \sysname proxies to the clients to minimize round-trips, and with and without compression. We use emulation to sweep through a large space of network configurations, and also present a smaller set of results using $6$ vantage points on the Internet.

To be able to control network characteristics tightly, we test those pages latencies to whose servers are negligible to begin with, as we can then easily evaluate the impact of additional latency. We thus deploy a \sysname proxy in Azure US WEST 2, and identify the $100$ most popular HTTP\footnote{We discuss HTTPS in \S\ref{subsec:https}.} sites reachable within $15$~ms latency from this location. The \sysname proxy uses $4$ cores and $16$~GB memory, and runs \texttt{PhantomJS}~\cite{phantomjsrelease}. 
For automating Web page loads and recording performance metrics, we use \texttt{sitespeed.io}~\cite{sitespeedio}. 

We measured several metrics, including: page load time, time to first paint, last visual change, speed index, and time for $85\%$ visual completion (viz$85$, indicative of when most of the visual content is populated). While we observed large differences in absolute values of these metrics, our interest is in \emph{percentage improvement} of such metrics when loading pages through \sysname compared to today's default. We find that metrics other than PLT and last visual change show results closely consistent with each other, but PLT and last visual change show somewhat smaller improvements. This is due to the dependence of PLT and last visual change on ad content (\eg when the network RTT is $160$~ms, enabling and disabling ad blocking changes median PLT by $8\%$, while there is no change in viz$85$), which requires multiple round-trips even with \sysname (due to ad requests being generated during rendering). Thus, the rest of our evaluation uses viz$85$.



\subsection{Performance on emulated networks}
\label{subsec:simulation}

We deploy the \sysname proxy and the client within the same Azure's US WEST 2 data center, thus ensuring that all latencies involved (proxy-server, client-proxy, and client-server) are negligible, and bandwidth can be controlled at the client. This allows us to vary network conditions and measure differences in page loads with and without \sysname. Page loads without \sysname are referred to as the default case. In keeping with our observation about being able to achieve close proximity between proxies and servers (\S\ref{sec:consolidation}), we always set the client-proxy latency to be the same as the client-server latency, with the proxy-server latency remaining as is (\ie without any manipulation.) The client runs the Chrome browser on Ubuntu virtual machines with $2$ cores and $8$~GB memory.

\begin{figure}
\begin{center}
\includegraphics[width=3in]{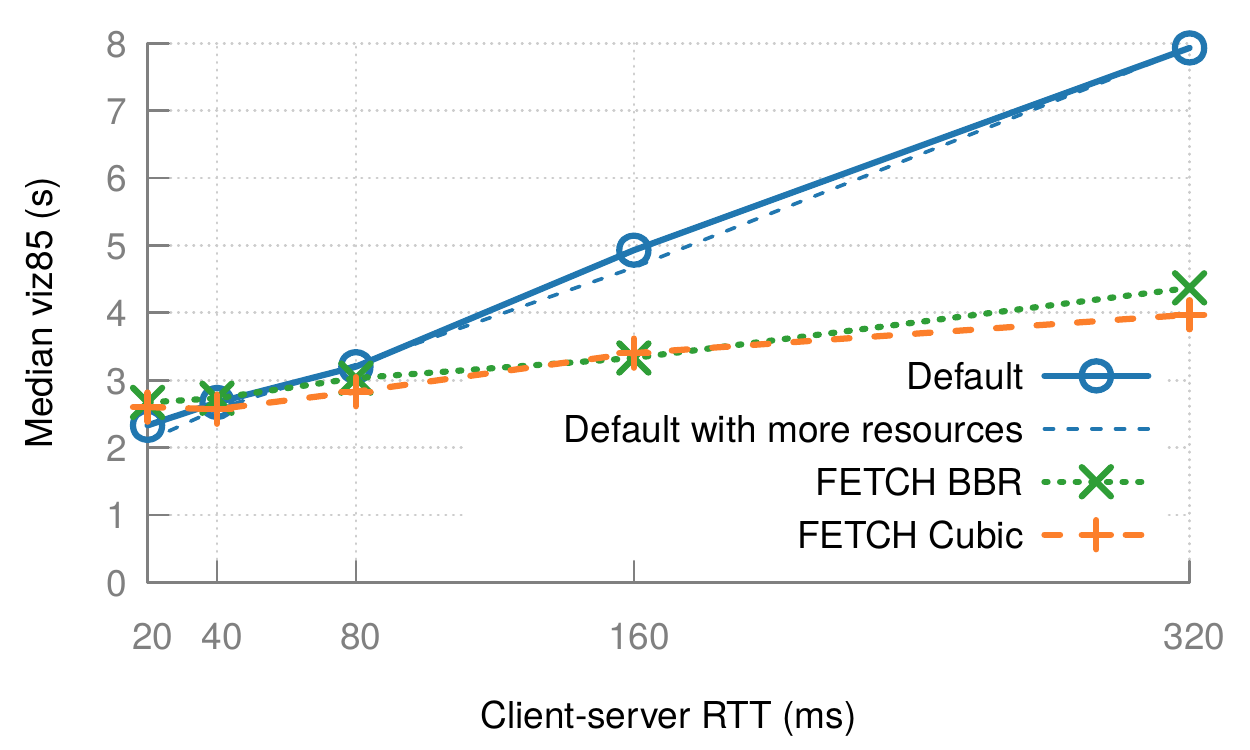}
\caption{\em With bandwidth fixed at $10$~Mbps, as client-server latency increases, the default case shows a very large linear increase in viz$85$ time. Using \sysname, the dependence on the network RTT is substantially reduced, resulting in a large improvement for clients with large latency to the server.}
\label{fig:results:increasingLatency}
\end{center}
\end{figure}

\parab{The impact of latency:}
Fig.~\ref{fig:results:increasingLatency} shows the median viz$85$ time across $100$ pages, with increasing client-server RTT, bandwidth fixed at $10$~Mbps, and with no (added) loss. Results for other bandwidth settings show similar trends. In the default case, as the RTT increases, viz$85$ time shows a large linear increase. In contrast, \sysname diminishes this dependence, showing much smaller increases in viz$85$ time. In the absence of significant loss, the two transport variants tested with \sysname achieve similar results. We find that bandwidth plays a much more limited role beyond a few Mbps (in line with expectations, hence results excluded). We also test the default from a host with more resources ($16$ cores and $64$~GB memory), but this does not reduce viz$85$ time substantially.

For clients with large latencies to Web servers, \sysname can provide a large speedup --- at $160$~ms latency to the Web server, \sysname loads pages $32\%$ ($1.6$~s) faster. Such large latencies are in fact typical in many parts of the World, as we discuss later in \S\ref{subsec:cdn_issues}.

\begin{figure}[t]
\begin{center}
\includegraphics[width=3in]{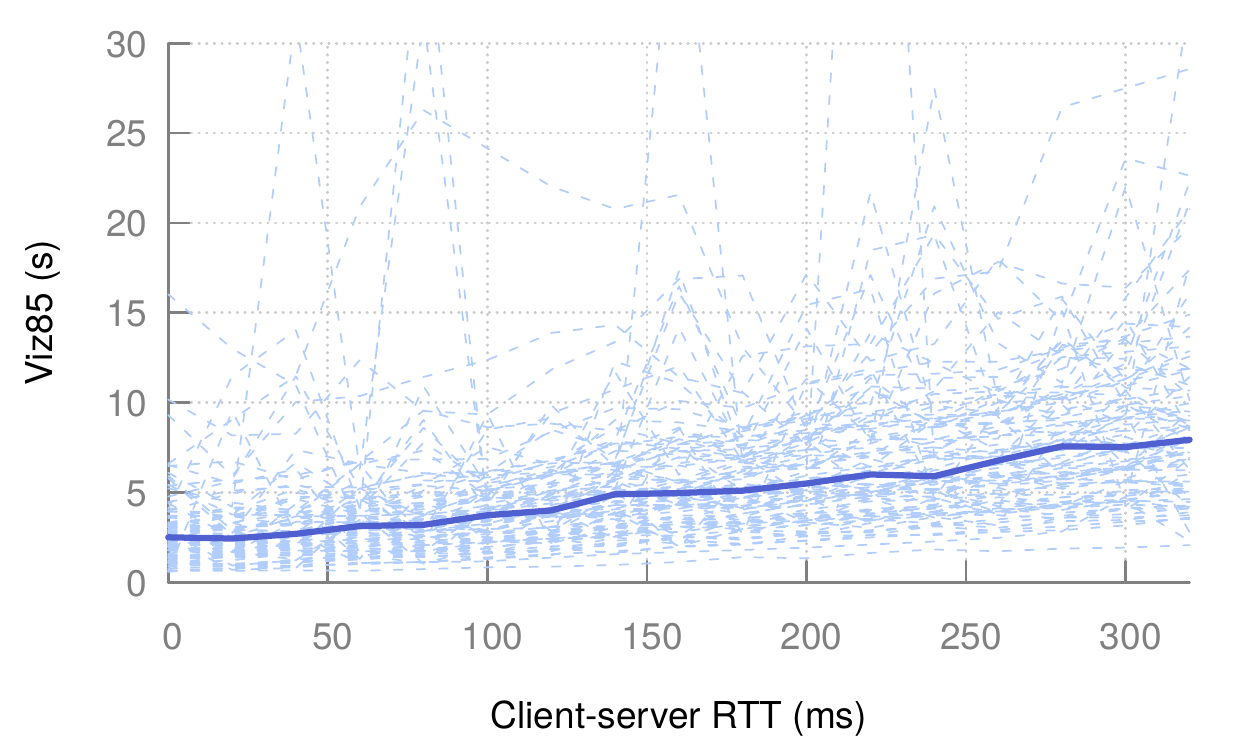}
\caption{\em As client-server RTT increases, load time increases linearly, with every $10$~ms of RTT increase, adding $186$~ms in the median. The individual dashed, light lines are for individual pages, and also show the linearity, albeit with variations for some pages, and with different slopes. The solid line represents the medians.}
\label{fig:rttDependence}
\vspace{10pt}
\end{center}
\end{figure}

An interesting by-product of our investigation is the dependence of page load times on client-server RTTs. The most frequently cited work on the impact of increasing RTT is Mike Belshe's $2010$ measurement of $25$ popular Web pages~\cite{belshe2010more}. Others have also quantified the relationship between measured \emph{last-mile} latency and page load times~\cite{sundaresanThesis, bishofLastmile} over small numbers of pages (less than $10$). We provide fresh measurements of this using our setup, which allows tight control of latency starting from nearly zero (under $10$~ms). Note that this kind of measurement is only made possible by our measurement observation in \S\ref{sec:consolidation}, with other setups starting from the already significant latencies they observe. While record-and-replay tools could also be used to produce such results, they often add significant inaccuracy (\eg $8\%$ in the median with MahiMahi~\cite{mahimahiDemo}). Fig.~\ref{fig:rttDependence} shows for each of the $100$ pages tested (which are all within $10$~ms from our client), how viz$85$ increases nearly in linear fashion with RTT (with bandwidth fixed to $10$~Mbps). The regression-based best-fit (over the medians at each RTT value, with RTTs being in seconds) is:
\begin{align}
t_{Default} & = 18.6 * RTT + 1.9
\label{eqn:default}
\end{align}

Thus, for every $10$~ms of increase in RTT, (median) load time increases by $186$~ms. (Faster compute affects the constant in that equation but not the linear factor.) Of course, there is substantial variation across pages, as shown by the individual lines in the plot, with some pages incurring substantially more RTTs than others. 

A similar regression-fit with \sysname results in:
\begin{align}
t_{\sysname}     & = 4.8 * RTT + 2.6
\label{eqn:oursystem}
\end{align}

Thus, \sysname's batching of results, together with optimized transport, results in a substantially smaller dependence on RTTs, although it incurs some additional overhead for the processing at the on-path proxy.

\begin{figure}
\begin{center}
\includegraphics[width=3in]{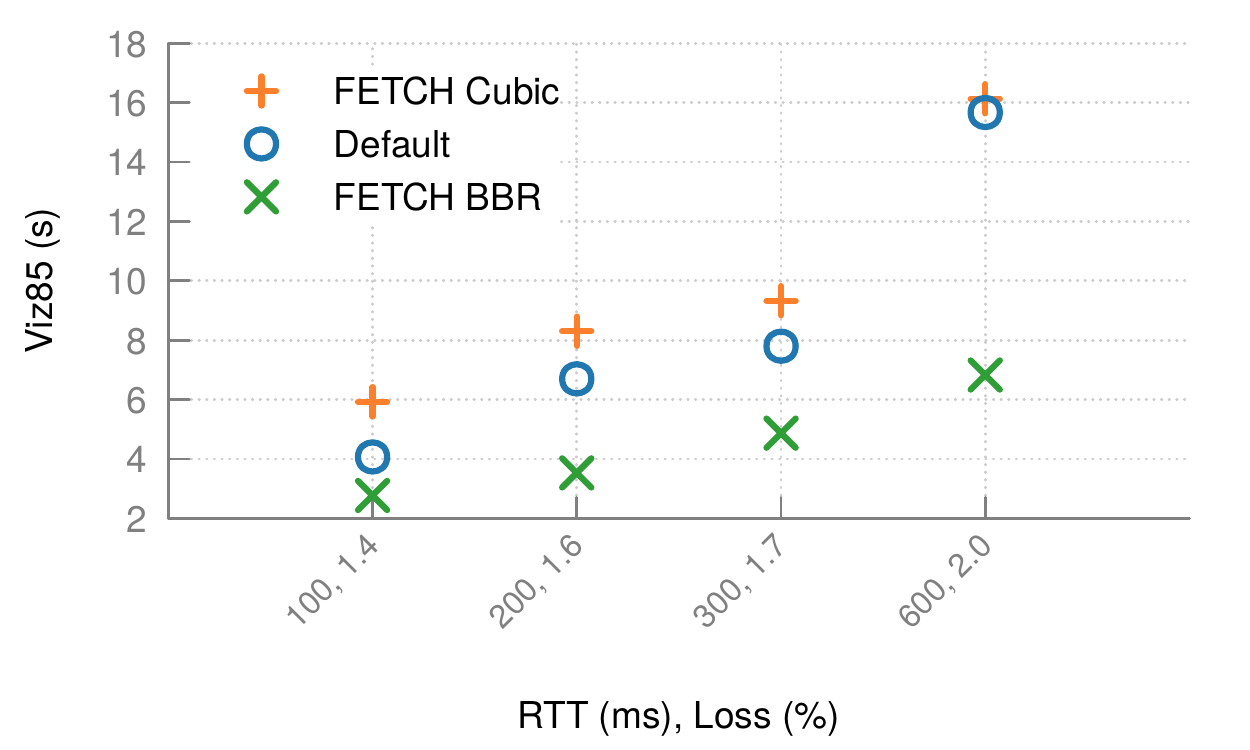}
\caption{\em With loss, \sysname performs worse than the default with TCP Cubic, but better with BBR between the \sysname proxy and the client.}
\label{fig:results:bernoulliloss}
\end{center}
\end{figure}

\parab{The impact of loss:}
Recent measurements from Google show that packet loss is also typically higher for connections with higher RTTs~\cite{quic}, so merely achieving improvements in settings with high RTTs but no loss is uninteresting. We thus simulate random network loss by 
traffic shaping at the client using Linux \texttt{tc}. For increasing RTTs, we additionally impose losses in line with Google's reported loss rates at those RTTs. 

Fig.~\ref{fig:results:bernoulliloss} shows median viz$85$ time across $100$ pages for different (RTT, loss $\%$) configurations. Here, using more sophisticated congestion control makes a large difference. With TCP Cubic as the congestion control mechanism between the server and the client, \sysname performs worse than the default as our current implementation uses a single TCP connection compared to the multiple connections for the default. Nevertheless, when BBR congestion control mechanism is used between the \sysname proxy and the client, \sysname performs substantially better -- with $100$~ms latency and $1.4\%$ loss, \sysname with BBR is $32\%$ faster than the default. For the other three configurations, these improvements are even larger.

\begin{figure}
\begin{center}
\includegraphics[width=3in]{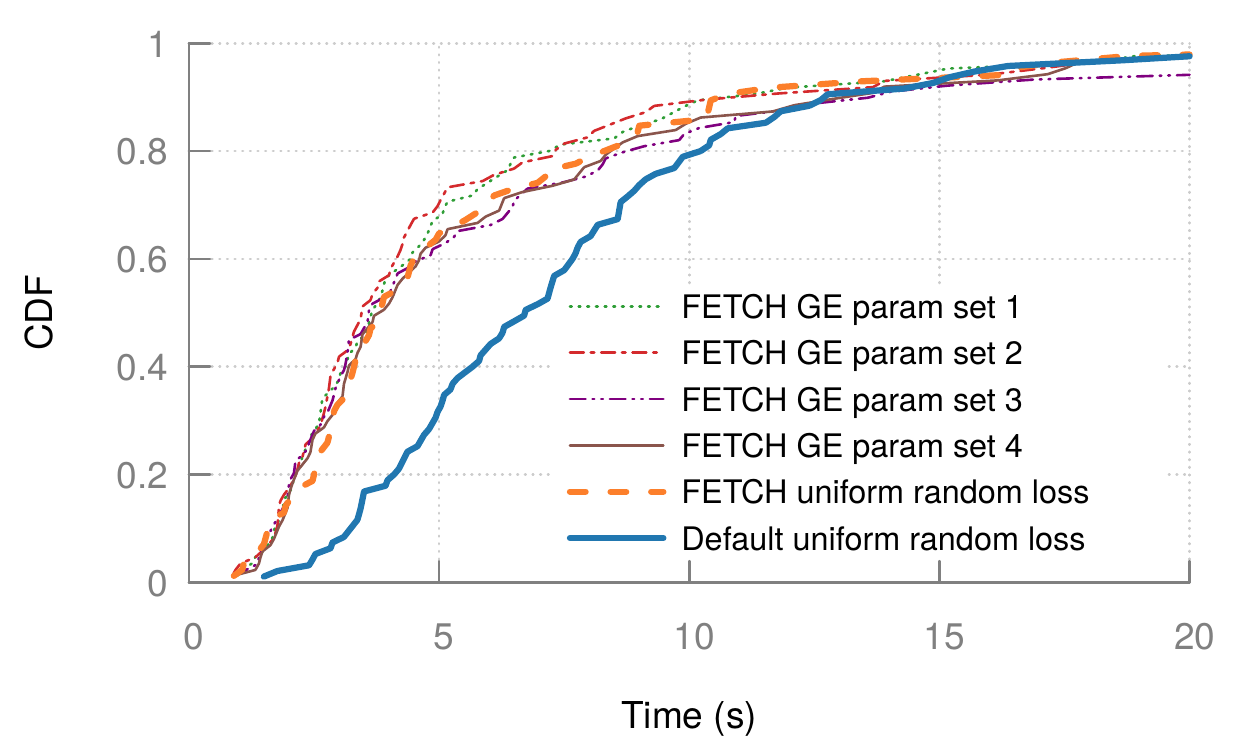}\vspace{-4pt}
\caption{\em Performance of \sysname with BBR congestion control under uniform random loss of $1.6\%$ and GE loss with different parameter sets which also correspond to $1.6\%$ loss. RTT is fixed to $200$~ms.}
\label{fig:results:gilbertelliottloss}\vspace{10pt}
\end{center}
\end{figure}
While the results in Fig.~\ref{fig:results:bernoulliloss} use a uniform random loss model, we also tested performance using a bursty Gilbert Elliott (GE) loss model~\cite{geloss}. In order to compare performance under uniform random loss and GE loss, we identify $4$ GE parameter sets which result in the same average loss rate of $1.6\%$. The GE model uses the parameters $p$, $r$, $1-h$ and $1-k$, which we set (in the same order) as follows for our $4$ configurations - set $1$: \{$1.096\%$, $50\%$, $70\%$, $0.1\%$\}, set $2$: \{$0.877\%$, $40\%$, $70\%$, $0.1\%$\}, set $3$: \{$0.658\%$, $30\%$, $70\%$, $0.1\%$\} and set $4$: \{$0.438\%$, $20\%$, $70\%$, $0.1\%$\}. Fig.~\ref{fig:results:gilbertelliottloss} shows no significant differences between the two loss models -- BBR is robust to both loss models. The rest of our experiments use BBR between the \sysname proxies and the clients.



\begin{figure}
\begin{center}
\includegraphics[width=3in]{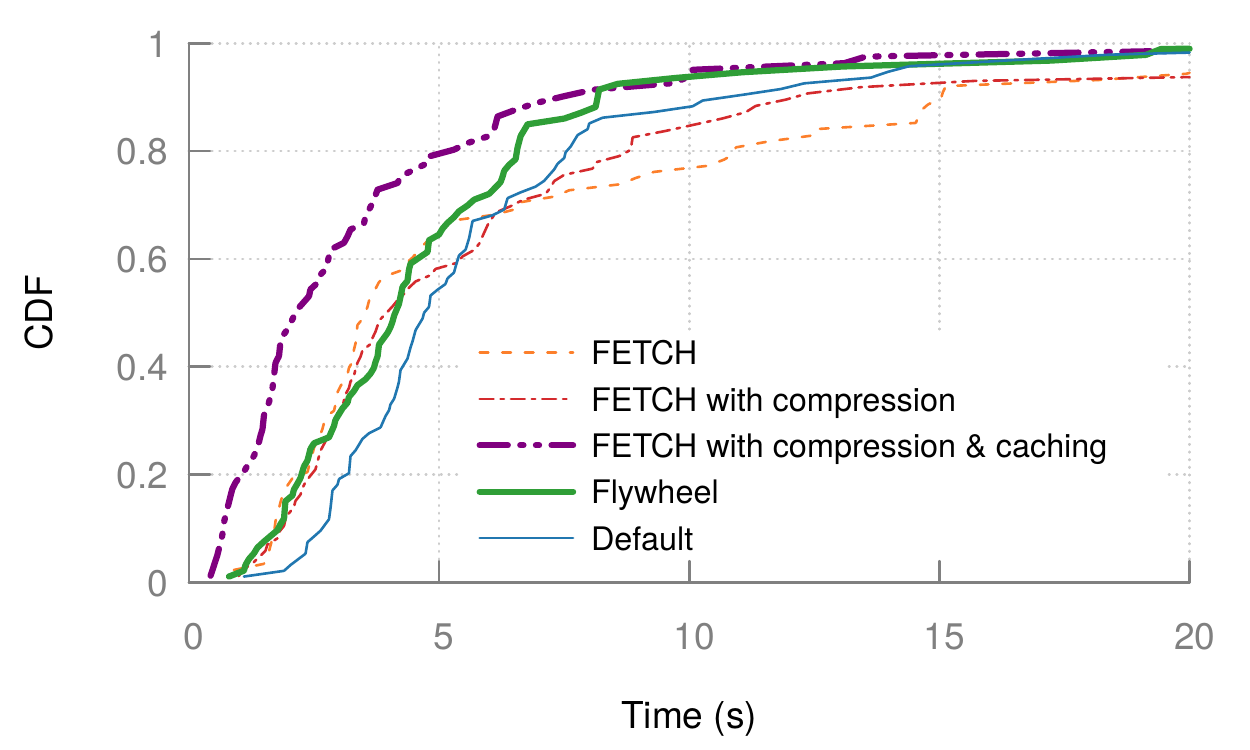}\vspace{-4pt}
\caption{\em Performance of \sysname is comparable to that of Flywheel in the median even without critical optimization like caching compressed content. A \sysname node which deploys both compression and caching is $48\%$ faster in the median compared to Google's Flywheel.}
\label{fig:results:flywheel}
\end{center}
\end{figure}

\parab{Compression and Flywheel:}
Google's Flywheel~\cite{agababov2015flywheel} uses a proxy-based approach to compress Web page resources with GZip and WebP as appropriate before transmitting them to clients. 
While Flywheel is targeted at mobile devices, the same functionality is also made available by Google for the desktop Chrome browser in the form of the ``Data Saver'' extension~\cite{datasaver}. 

We evaluate the same $100$ Web pages with Flywheel, with \sysname, and with neither. For \sysname, we test two additional variants: (a) with WebP compression for images using the webp-imageio library~\cite{imageio} and GZip compression for text content \emph{on the fly}; and (b) with compression and caching at the \sysname proxy, such that cached resources are not compressed on the fly. Flywheel, being a compression proxy, is most useful in low-bandwidth settings, so we first discuss results for a client with $2$~Mbps bandwidth (and a fixed network RTT of $160$~ms).

To make the effect of compression comparable, we first loaded several pages with and without Flywheel, storing their default images and Flywheel's WebP versions of the same images, and set parameters for our compression to match Flywheel's resulting images. For the \sysname variant with caching, we use a hashmap based in-memory key-value store to cache the compressed content. In order to analyze the effect of caching, we load each page with this \sysname node twice (clearing the client-side cache, but populating the \sysname cache) and consider every second page load. (We realize that this is the best-case caching result; but running the comparison over popular pages implies that Flywheel also gets the same or similar benefit, as they also cache compressed versions of resources for popular pages~\cite{agababov2015flywheel}.)

As Fig.~\ref{fig:results:flywheel} shows, Flywheel's compression makes it faster than the default case in this bandwidth-constrained setting. \sysname without compression is $16\%$ faster than Flywheel in the median, but is slower at higher percentiles. At this constrained bandwidth, the lack of compression hurts \sysname's load times, particularly for large image-rich pages. Interestingly, adding on-the-fly compression actually makes \sysname slower in the median, because this incurs significant compute time on the critical path, eliminating the advantage from the faster transfer of compressed resources. As expected, the \sysname variant with both compression and caching is substantially faster: $48\%$ faster in the median compared to Google's Flywheel. In high-bandwidth settings, even \sysname without compression and caching is faster than Flywheel -- at $10$~Mbps, $29\%$ faster in the median (plot omitted). 

Note that Flywheel benefits from Google's extensive data center and networking infrastructure; and as is typical across proxy-based designs so far, routes client requests to locations nearest to them. In contrast, \sysname only depends on a small number of public cloud virtual machines. Achieving proximity to a large number of clients (for near-client proxies) is fundamentally more challenging than achieving proximity to most Web servers (like for \sysname). 

\subsection{Performance across the Internet}
\label{subsec:realexp}

\begin{figure*}[t]
  \centering
  \subfigure[]{\label{fig:usa}
    \includegraphics[width=2.1in]{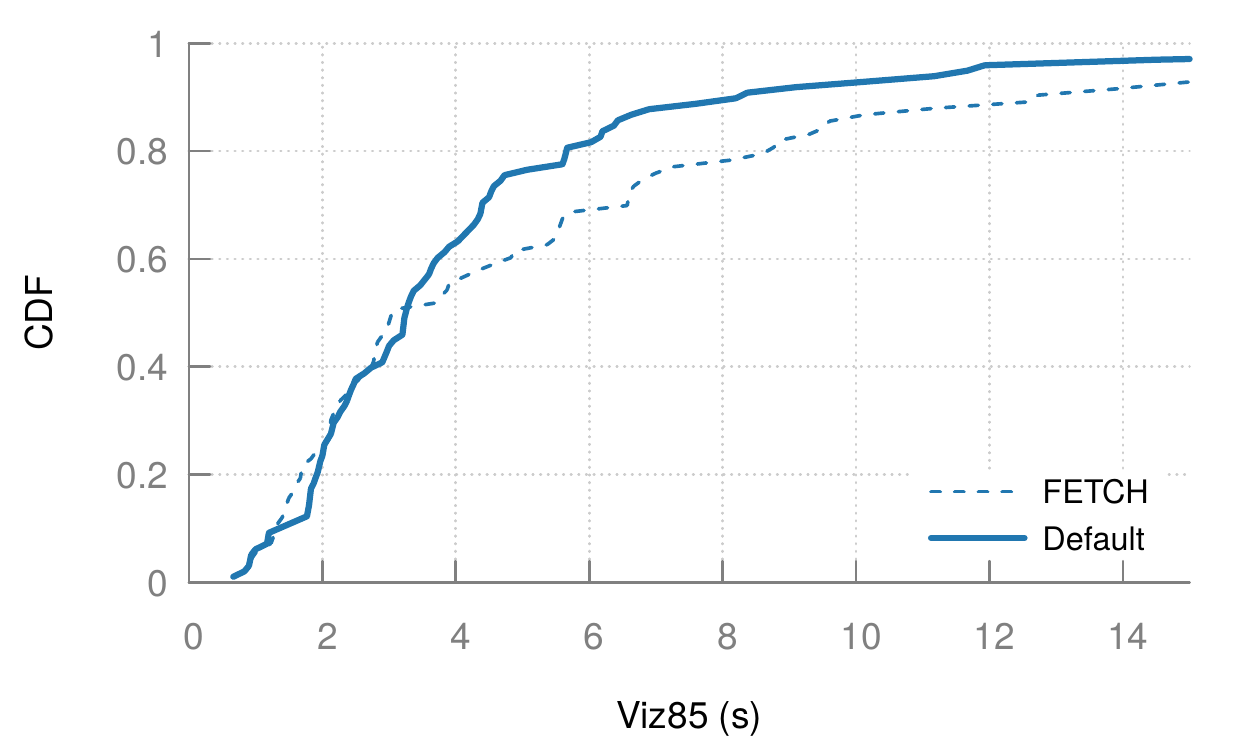}}
  \subfigure[]{\label{fig:eu}
    \includegraphics[width=2.1in]{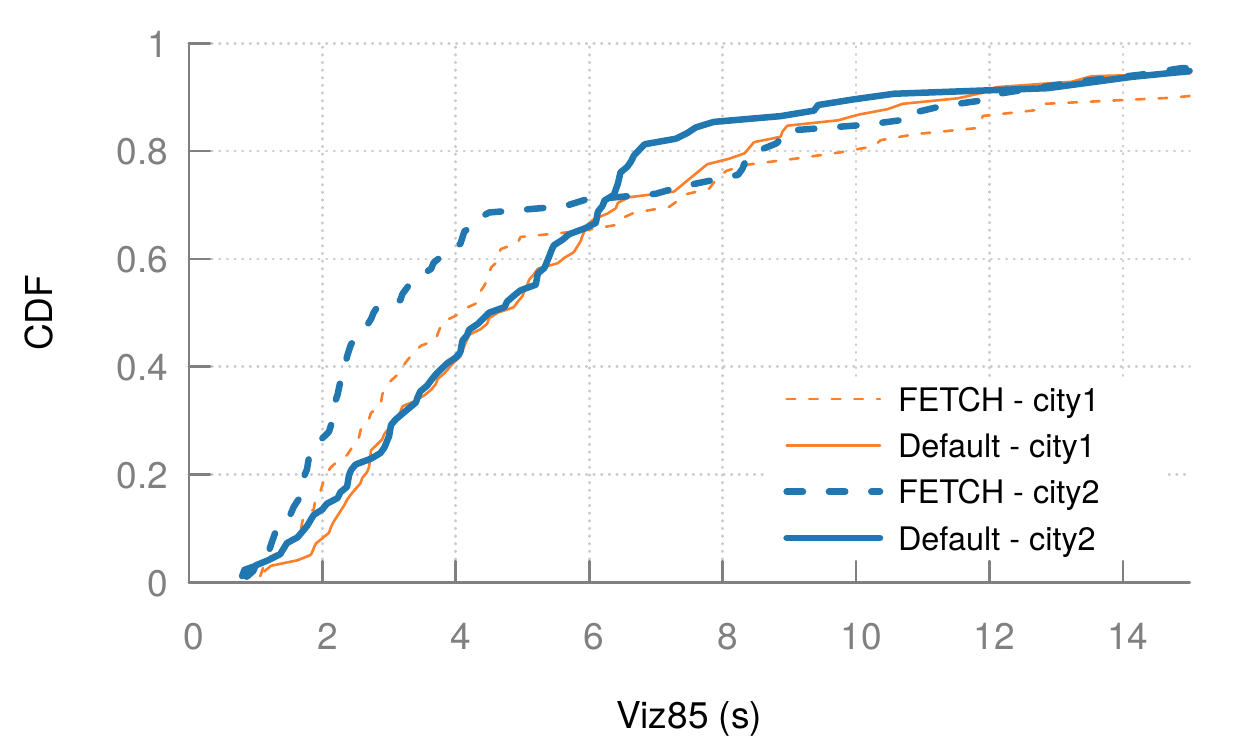}}
  \subfigure[]{\label{fig:asia}
    \includegraphics[width=2.1in]{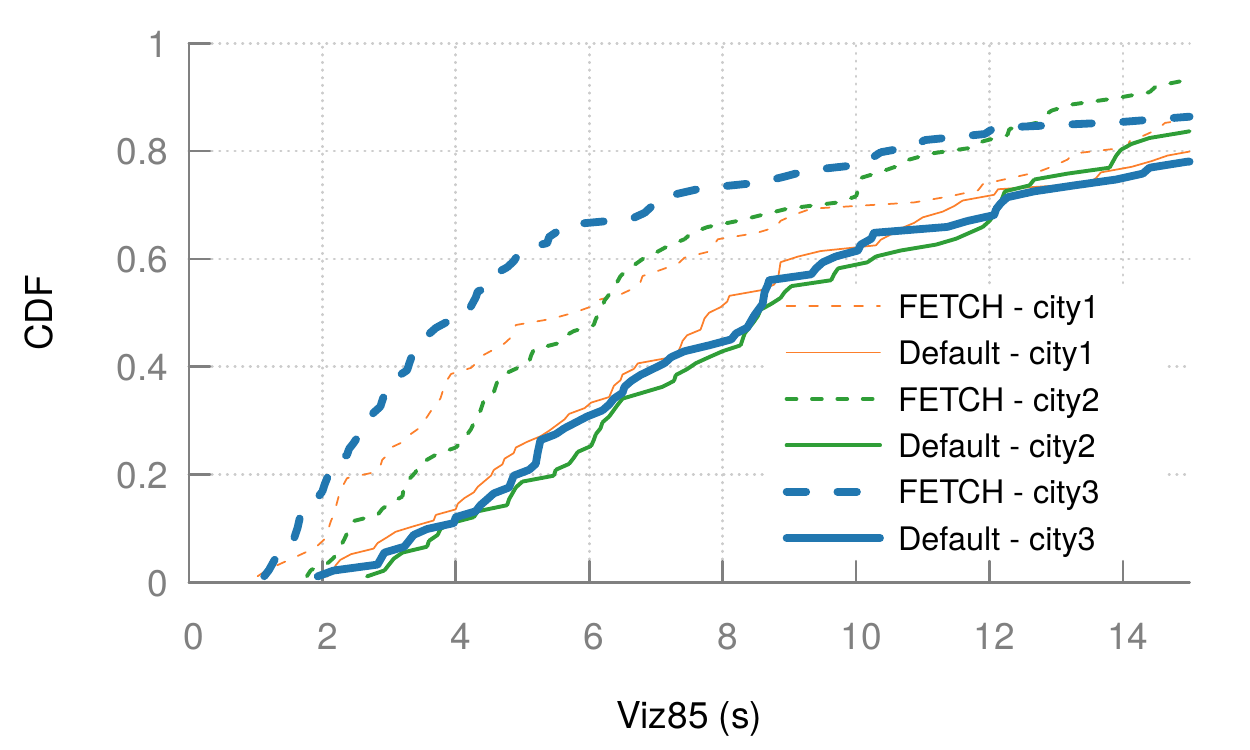}}
  \caption{\em \sysname vs. default, for: (a) a client in the US, (b) $2$ clients in Europe; and (c) $3$ clients in Asia.}
   \vspace{10pt}
  \label{fig:real_viz85}
\end{figure*}

Apart from our sweep across many emulated network configurations, we also include results on a small evaluation across real Internet connections. 

We run experiments from $6$ vantage points access to which was obtained informally through friends. To side-step privacy concerns, we still test the same $100$ pages as in \S\ref{subsec:simulation}, instead of using real browsing data. 
Our $6$ clients load the same $100$ pages with and without \sysname across their varied network connectivity. Fig.~\ref{fig:real_viz85} shows the CDFs of the viz$85$ times from the different locations.
For each of the $3$ clients in Asia (Fig.~\ref{fig:asia}), due to their high latency to the Web servers, \sysname clearly outperforms the default, with median improvements between $26\%$ and $50\%$. 
For the well-connected clients in the US (Fig.~\ref{fig:usa}) and Europe (Fig.~\ref{fig:eu}), \sysname is comparable or better than the default in the median, but slower at higher percentiles.

\begin{figure}
\begin{center}
\includegraphics[width=3in]{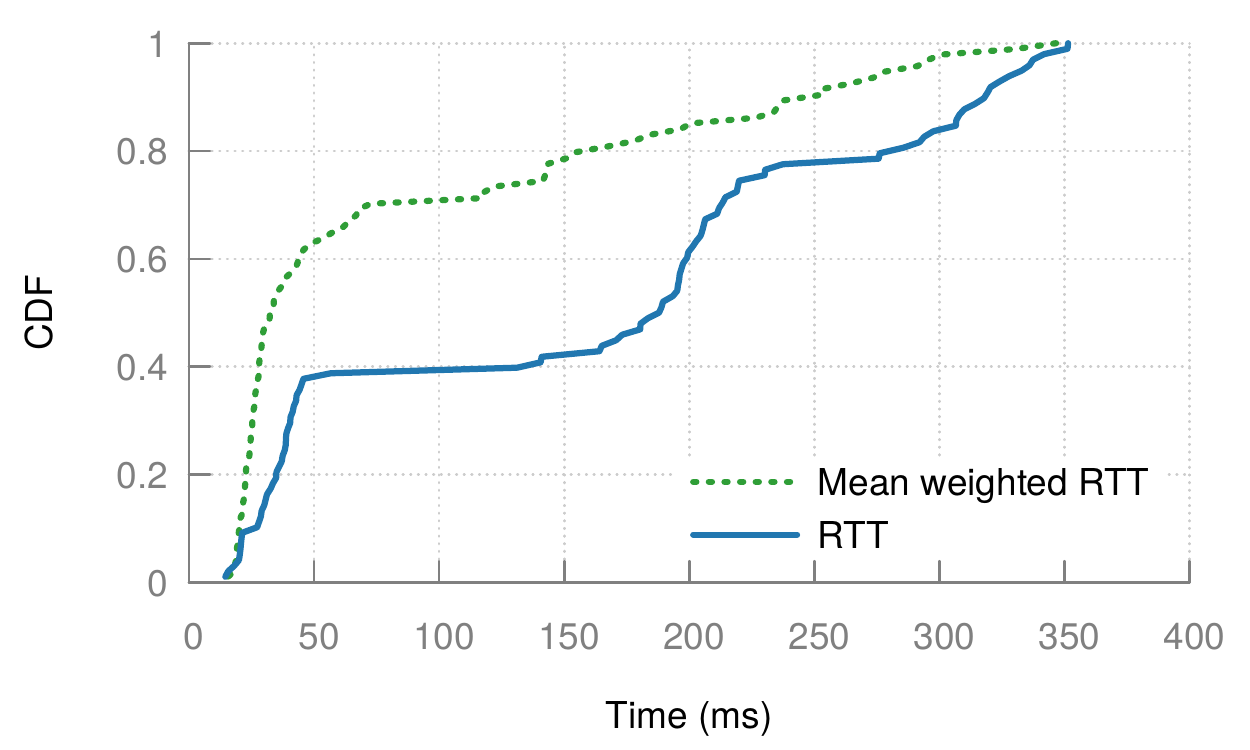}
\caption{\em Although the median RTT in city$2$ of Fig.~\ref{fig:eu} to the $100$ Web services is $188$~ms, the mwRTT is still low: $32$~ms in the median.}
\label{fig:results:zurich_mwrtt}
\end{center}
\end{figure}

To understand performance differences at higher percentiles, we examined RTTs and mwRTTs (like in \S\ref{subsec:thirdpartydomains}) for one European client. Fig.~\ref{fig:results:zurich_mwrtt} shows the results. While the median RTT is indeed large (as most of these Web servers are in the US) at $188$~ms, the mwRTT is only $32$~ms, with $\sim$$40\%$ of bytes served from replicas somewhere closer than the US. Thus, many resources in this setting are fetched over small network RTTs, negating the advantage of \sysname, which is larger when more data is being fetched over high-latency networks. As noted in \S\ref{subsec:simulation}, large populations of clients are in such regimes (\eg with high last-mile latency).

\section{Competitiveness with a CDN?}
\label{subsec:cgnCDN}

Content distribution networks are a well-established way of speeding up Web services today. While CDN usage is far from ubiquitous, we would nevertheless like to understand how Web page acceleration with \sysname would compare to CDN-based acceleration.


\subsection{The limitations of CDNs}
\label{subsec:cdn_issues}

Content distribution networks build worldwide infrastructures to establish proximity between content and users, and thus allow users to reach their services at lower latencies. However, they are far from a catch-all solution for several reasons:

\parab{Content coverage:} Due to their expense, CDNs are estimated to be used by under $10\%$ of the top $32$,$000$ most popular Web sites, with usage dropping with decline in popularity~\cite{gilad2016cdn}. Further, most CDN-enabled Web sites today do not use CDNs for full-site delivery -- often, the initial HTML, as well as any dynamic or personalized content must still be fetched from the content originator's servers, which could incur a large latency. 

\parab{Last-mile latency:} Last-mile latencies in many parts of the World are extremely poor. For instance, in Pakistan, last-mile latencies for $3$ out of $5$ measured providers are around $100$~ms in the median, and average latency to M-Lab servers within the same city exceeds $100$~ms~\cite{awan2015measuring}. While, not strictly last-mile, median latencies to the nearest M-Lab servers are $\sim$$200$~ms in Africa, Asia, South America, and Oceania~\cite{Hoiland-Jorgensen} even though there are several M-Lab servers in most of these geographies. In Asia, even the fastest of $4$ measured CDNs required an average of $1$~sec to deliver a small (latency-bounded) $12$~KB object, compared to under $0.4$~sec in Europe and North America~\cite{CDNcompare}. On mobile networks, last-mile latencies will likely continue to be poor in many parts of the World for several years, as the HSPA/HSPA+ standards have been or are being deployed, with latencies on the order of $100$~ms~\cite{ilyaHSPA}. For users with such large last-mile latencies, even latency to \emph{geographically} nearby CDN servers can exceed $100$ milliseconds, and the many RTTs needed to fetch Web content from these servers can still add up to large page load times and poor user experience. 

\begin{figure}
\begin{center}
\includegraphics[width=2.8in,clip]{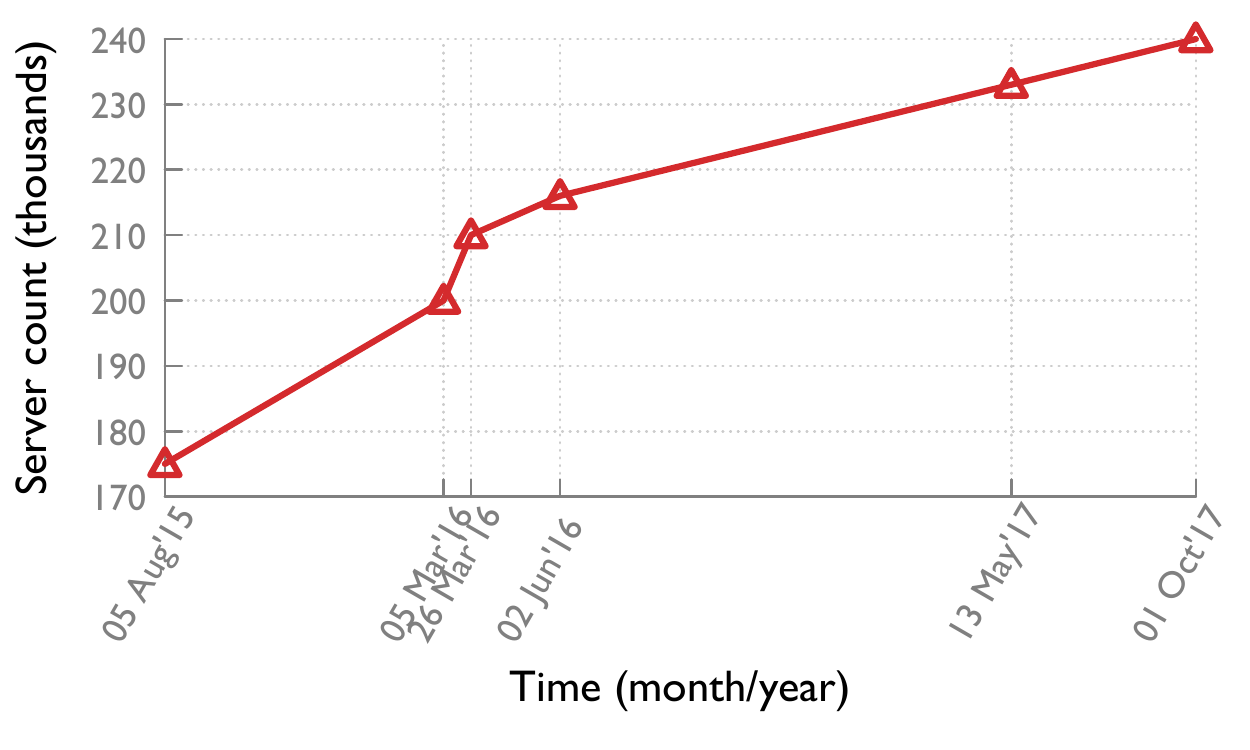}
\caption{\em Akamai's extensive, constantly growing footprint.}
\vspace{-4pt}
\label{fig:akamaiGrowth}
\end{center}
\end{figure}

\parab{Complex, expensive infrastructure:} CDNs depend on a large, expensive, and constantly spreading infrastructural footprint. While such networks do not release information about their growth over time, we were able to gather some (approximate) data points using the ``Internet Wayback Machine''~\cite{waybackMachine} on Akamai's public page disclosing their network's size~\cite{akamaiInfra}. As shown in Fig.~\ref{fig:akamaiGrowth}, on average, Akamai adds $\sim$$2$,$500$ servers~/~month. In the same period, the number of networks to which they connect increased from $1300$$+$ to $1600$$+$. Relatedly, provisioning CDN infrastructure is a challenging problem because of the need to forecast demands to deploy hardware. When provisioning fails to handle content demand, CDNs must redirect users to farther away servers. 

\greybox{
The above challenges stem from a fundamental issue: establishing low-latency connectivity covering most users is made difficult and expensive by the spread of users and the diversity of their last-mile connectivity. In contrast, as our measurements show (\S\ref{sec:consolidation}), achieving proximity to most Web servers is extremely easy and cheap.}



\begin{figure}
\begin{center}
\includegraphics[width=3in]{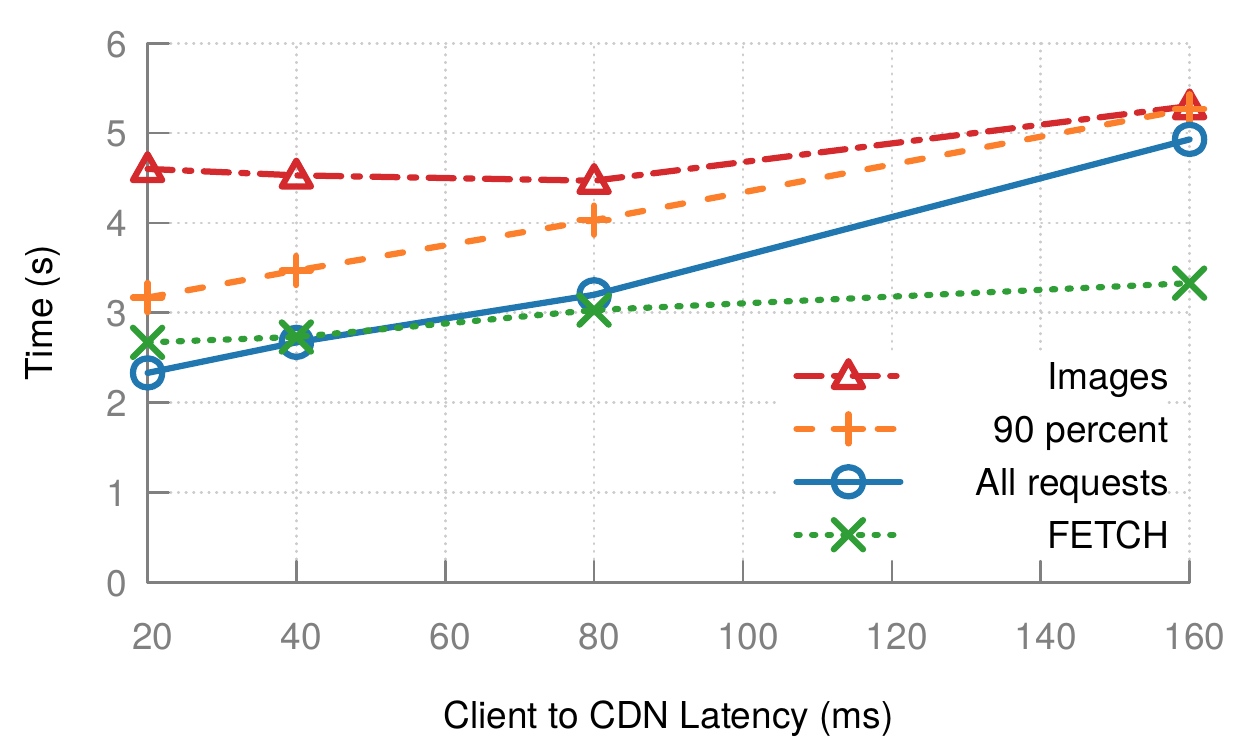}
\caption{\em With client-server latency set to $160$~ms, we emulate results for $3$ possible uses of a CDN, under increasing client-CDN latency: with the CDN only serving all images; serving a random $90\%$ of requests; and serving all requests.}
\label{fig:cdn_model}
\end{center}
\end{figure}
\subsection{CDN vs. \sysname acceleration}
\label{subsec:cgn_cdn_result}

Our results in Fig.~\ref{fig:results:increasingLatency} comparing \sysname-based Web page delivery to the default show substantial performance improvements. However, these results were obtained by adding additional latency to all requests from a client to emulate large client-server latency. In a CDN-supported Web service, this is not a realistic comparison  -- CDNs make extensive efforts to place their servers closer to clients, and resources fetched from CDNs may thus be delivered over much smaller RTTs. This is already evident from our results in Fig.~\ref{fig:real_viz85} and Fig.~\ref{fig:results:zurich_mwrtt}, with the latter showing the median mwRTT to be significantly smaller than the median RTT to Web servers. Hence, we next model how fetching Web pages using \sysname compares to fetching pages supported by a CDN.

We emulate different extents of CDN support, with the CDN used for: (a) fetching only images\footnote{This is the byte-dominant part of most Web pages -- for Flywheel's deployment, images comprised $74\%$ of the bytes in pages served~\cite{agababov2015flywheel}.}; (b) fulfilling a random $90\%$ of requests; and (c) fulfilling all requests. The rationale for evaluating options (a) and (b) stems from our conversations with a large CDN operator, which indicate that most of their customers do not use their services for full-site delivery, rather offloading only parts of it. For these experiments, we add lower amounts to latency to requests fulfilled through the CDN than for other requests, thus modeling smaller client-CDN latency and a larger client-server latency.

Results in Fig.~\ref{fig:cdn_model} show the dependence of $85\%$ visual completion time on the client-CDN latency (with client-server latency fixed to $160$~ms). As expected, the more the content served through the CDN, the better the performance. However, as the client-CDN latency increases, performance degrades quickly. In contrast, as observed earlier, \sysname effectively reduces the dependence on the network RTT, thus gaining an advantage when RTTs are larger. Thus, particularly for users with large last-mile latency, \sysname would continue to have a performance advantage over CDN-based delivery. 

A big caveat to these results is that they do not account for the possibility that CDNs, like \sysname, themselves may push towards delivering batched data in minimal RTTs. CDNs do already use more aggressive transport than most Web servers, with some using congestion window sizes as large as $46$ packets~\cite{cdnInitWindow} compared to the Linux default of $10$. Likewise, Akamai is already starting to use QUIC~\cite{akamaiQUIC}.
It should be rather obvious that \emph{if} a CDN performs full-site delivery, and uses the \emph{same} optimizations as a proxy (for both networking and resource prioritization), then it will outperform the proxy, by virtue of incurring a shorter RTT. 
In the following, we use our experimental data to build a simple model to explore a scenario where CDNs use the same aggressive optimizations towards batching and better transport as \sysname.


We construct a simple, linear model of how the involved round-trip latencies --- client-server ($RTT_s$), client-\sysname ($RTT_g$) and client-CDN ($RTT_{lm}$) --- impact Web performance using CDNs and the \sysname. For simplification, drawing on our measurements, we assume that \sysname-server latency is nearly zero, \ie $RTT_g = RTT_s$. We also assume that enough bandwidth is available -- in our experiments, beyond a few Mbps, bandwidth changes only had a small impact on performance. We use the $10$~Mbps data to build our model.

The regression-based best fits for viz$85$ for default page loads ($t_{D}$)  and \sysname-supported page loads ($t_{\sysname}$) were already discussed in \S\ref{subsec:simulation} (Eqn.~\ref{eqn:default} and \ref{eqn:oursystem}).
Thus, \sysname depends much less on network RTT (less than $5\times$) than the default ($18\times$). However, the addition of an on-path proxy incurs some additional overhead in the form of the larger constant for \sysname ($2.6$~sec vs. $1.9$ for default). Based on this network vs. compute breakdown, we can construct two hypotheticals -- a well-tuned \sysname with near-zero processing overhead (\sysnameNoSpace\textsuperscript{*}), and a well-tuned CDN with \sysname-like dependence on the network (CDN\textsuperscript{*}) as follows:
\begin{align}
t_{CDN^*}   & = 4.8 * RTT_{lm} + 1.9   \\
t_{\sysname^*}   & = 4.8 * RTT_{s} + 1.9
\end{align}

\begin{figure}[t]
\begin{center}
\includegraphics[width=3in]{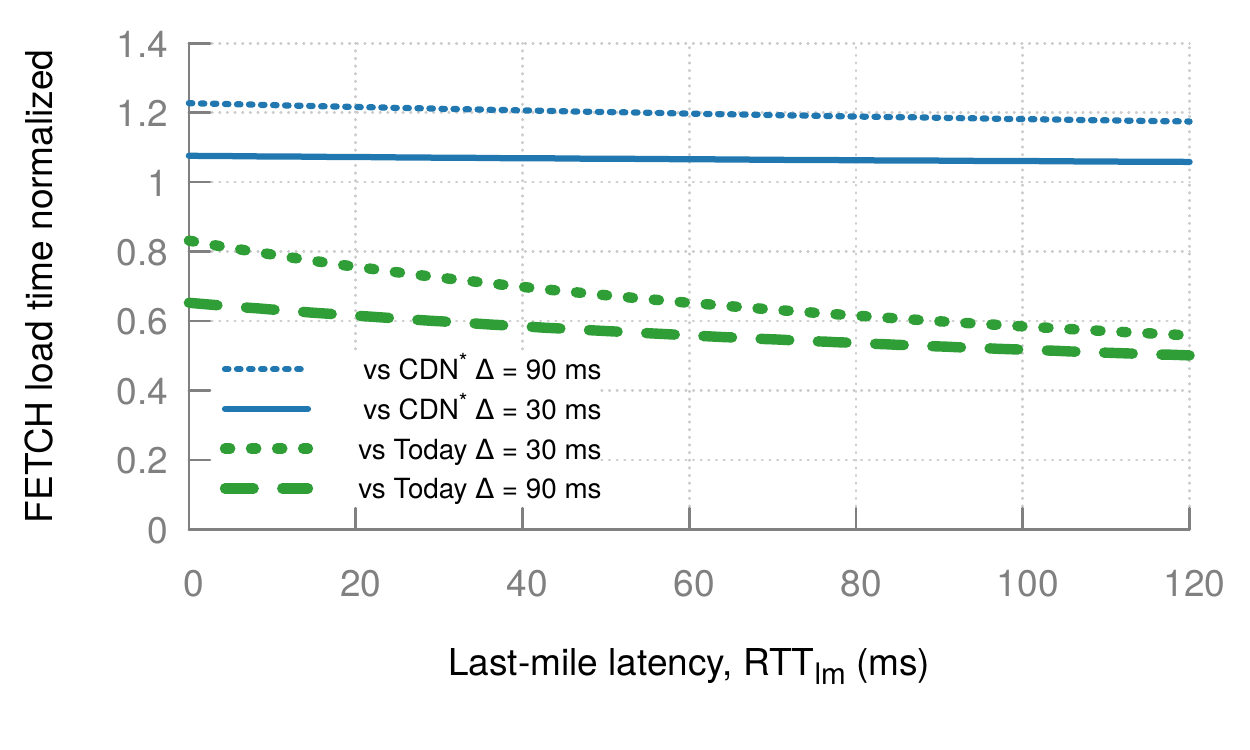}
\vspace{-12pt}
\caption{\em As last-mile latency dominates, \sysname's relative performance improves.\vspace{10pt}}
\label{fig:cdnVcgnIdeal}
\end{center}
\end{figure}

\sysnameNoSpace\textsuperscript{*} incurs $RTT_{s} > RTT_{lm}$, where $RTT_{lm}$ is the last-mile latency, which the CDN incurs. Suppose $RTT_{s} = \Delta + RTT_{lm}$ for some $\Delta > 0$. We can then estimate the performance of the CDN and \sysname using this model across different values of $RTT_{lm}$ and $\Delta$. Fig.~\ref{fig:cdnVcgnIdeal} shows these estimates for \sysname (normalized to performance as measured from the Web ecosystem today), and for \sysnameNoSpace\textsuperscript{*} (normalized to estimated performance for CDN\textsuperscript{*}). As our experimental results already showed, \sysname achieves substantial performance benefits over default page fetches today (with normalized performance values under $1$). Further, this advantage improves as last mile latency increases. 

Of greater interest here, is the comparison to an ideal CDN\textsuperscript{*}. As noted above, \sysnameNoSpace\textsuperscript{*} will achieve performance lower than CDN\textsuperscript{*}, but the difference is not very large. In particular, for small $\Delta$, performance is roughly $5\%$ worse than CDN\textsuperscript{*}. Note that a small $\Delta$ can be easy to achieve with relatively little replication (effectively, ``CDN-lite''), because it does not require presence at the edge. (Recall that $\Delta$ is the additional latency to a \sysname proxy \emph{beyond} the last-mile.) Further, performance relative to CDN\textsuperscript{*} improves (\ie becomes closer to $1$ in the Fig.~\ref{fig:cdnVcgnIdeal}) as the last-mile latency increases -- in such regimes, the advantage of CDN\textsuperscript{*} is reduced by the last-mile latency forming a larger fraction of the client-server RTT. Lastly, note that this model assumes the CDN does not incur any round-trips to the server; which is unlikely to be the case for dynamic, personalized content. Thus, a simple design based on a small number of worldwide sites can  not only compete effectively with today's Web ecosystem, but also come within striking distance of a well-tuned CDN doing full-site delivery and aggressive network optimization.



%
\section{Towards realizing \sysname} 
\label{discussions}

While we already have a prototype implementation, it is worth discussing what it would take to deploy \sysname and make it widely available.

\subsection{Wouldn't \sysname be very expensive?} 
\label{subsec:cost}

\begin{figure}[t]
\begin{center}
\includegraphics[width=3in]{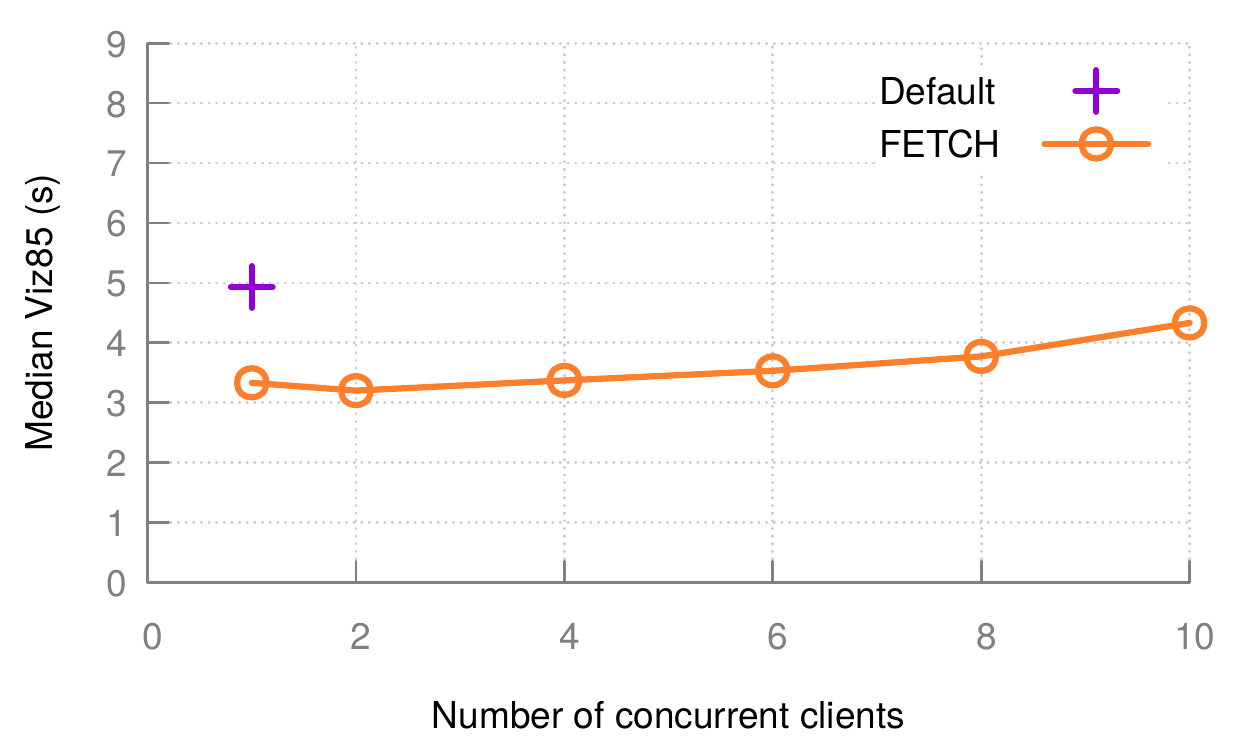}
\caption{\em A small number of concurrent clients do not substantially deteriorate a \sysname instance's performance. Client-\sysname latency = $160$~ms; bandwidth = $10$~Mbps.}
\label{fig:parallelization}
\end{center}
\end{figure}

We present a coarse per-user cost analysis based on our (first-cut) \sysname proxy implementation running on the same hardware as for our experiments in \S\ref{sec:results}.

The average page size across our $100$ test pages is $\sim$$2$~MB, and requires $5.2$~seconds of service time at the \sysname proxy, computed as the time from when the proxy receives a request to when it finishes processing it. The costs (averaged across Azure sites for the \texttt{D4s v3} instances we used) of compute and networking are $\$0.431$ per hour and $\$0.087$ per GB~\cite{Azurepricing}. Using an estimate of $3$,$000$ Web pages per month fetched per user~\cite{theNielsenReport}, the monthly costs of network traffic and compute per user are \$$0.52$ and \$$1.86$ respectively, coming to a total of $\$2.38$ per user per month.

Further, a single \texttt{D4s v3} instance can concurrently support several users' requests without much performance degradation. We vary the numbers of clients fetching the same set of $100$ Web pages simultaneously through a single instance, without any caching. Each client also shuffles the order in which it fetches the set of Web pages to avoid server-side overheads from (near) simultaneous arrival of these requests. As Fig.~\ref{fig:parallelization} shows, performance does not deteriorate for up to $4$ clients. Adjusting the compute costs (the network component remaining unchanged) for $4$ concurrent clients brings the total cost per user per month down to under $\$1$.
Admittedly, this cost estimate assumes $100\%$ system utilization, while in practice demand can be expected to be variable over time, necessarily leading to some inefficiencies. However, in the cloud environment, it is fairly trivial to adjust compute resources to handle demand fluctuations. 

Our estimate is based on an unoptimized proxy loading pages to completion. Recent work has already demonstrated that such proxies can be made more efficient, \eg by not performing rendering~\cite{shandian}, and not executing all scripts~\cite{proxyEfficient}. Thus, an operational \sysname system could likely be run at much lower cost.

\subsection{Who runs \sysname?}
\label{subsec:whoDeploys}

The lowest-hanging fruit is to use the measurement observation behind \sysname's design in existing proxy-based systems. For instance, Google's Flywheel~\cite{agababov2015flywheel} and Facebook's Free Basics~\cite{freeBasics} could be made more efficient simply by optimizing the placement of the proxies involved nearer to the servers. For instance, for Free Basics, recent work~\cite{sen2017inside} reveals that each Free Basics Web request uses (in addition to a proxy near the client) a proxy that interacts with the target Web service. This proxy is located in one of two Facebook data centers (Lule\r{a}, Sweden and Prineville, USA). By using a small number of proxies in the right locations, together with simple known transport optimizations, Web performance in such deployments could be improved substantially.

Beyond the above simplest deployment model, \sysname could essentially be operated by any independent third party. For instance, a browser extension could incorporate the client-side logic, and operate the proxies. The cost could be recouped via ads or a subscription fee for users\footnote{Note that $\$1$ per month is many times smaller than the price differential between bandwidth tiers in most of the World.}. This is operationally similar to operating a VPN service (except here the proxies do not merely forward traffic). In principle, even individual users could operate such a service for themselves. At present, running (sparsely used) proxies full-time at several locations would be cost prohibitive, but as cloud billing aligns closer to actual usage, this could become plausible.

While a low-expense or ad-based service that speeds up the Web could be attractive to many users, this line of thought moves the burden of improving Web performance from Web service providers to users. This is potentially necessary to address the issue of slow-evolving Web service providers with inefficient services, which frustrate users. Another possibility, for service providers looking for cheaper alternatives to CDNs, is to make entry into the \sysname mappings paid. With this model, \sysname only accelerates Web services that incur a (potentially per-request) fee for being served over \sysname. Effectively, this involves deploying your Web service in a major cloud data center (or replicating in a small number, per \S\ref{subsec:cgn_cdn_result}) and then outsourcing your reverse proxy for \sysname to run.

\subsection{Security \& trust}
\label{subsec:https}

While our present prototype has only been tested with HTTP, the Web is rapidly moving to HTTPS. This presents a challenge for all proxy-based architectures, which fundamentally operate using network middlemen. In fact, even with CDNs, trust is not truly end-to-end; rather, in a majority of deployments, Web service operators trust CDNs with their private keys~\cite{cdnKeys}. \sysname can make a similar compromise, but at the client side, with the client having to trust \sysname -- the client uses a secure connection to \sysname, which uses a new secure connection to Web services. This is precisely how Free Basics supports HTTPS today~\cite{fbhttps}.


In the long-term, this solution is unsatisfactory. One could further ensure that client data is not seen even by a \sysname operator, using solutions like mbTLS~\cite{mbTLS}, running over private computing platforms like Intel SGX~\cite{intelSGX}.

\subsection{Further improvements}

We have only explored a deliberately simple design for \sysname here, but there is significant potential for further improvement. For instance, in computing the mapping between Web services and \sysname proxies, we always pick only the strictly lowest-latency pairings. This can reduce \sysname's benefits in some scenarios. Consider an example service $S$ that is replicated in two locations, one in Europe ($X_{E}$) and another in the US ($X_{U}$). Even if the $\sysname_{E}$-$X_{E}$ latency is larger than $\sysname_{U}$-$X_{U}$ latency by under a millisecond, European clients of \sysname will connect to $X_{U}$ through $\sysname_{U}$. This can be addressed by including multiple candidate proxies for each service together with their measured latencies in the mappings shared with clients, with client-side software deciding which mappings to use. For a client's most frequently visited services, it should even be possible to learn which \sysname proxy ultimately leads to the best performance. We also expect large improvements from minimizing processing time at the \sysname proxies.

There are also potential provider-side issues that could arise, \eg \sysname can redirect traffic for replicated services to one location (or a small number of locations with the above described improvement), causing load-balancing problems. At minimum, large providers already using such replication can opt-out of their service being proxied through \sysname.

\vspace{0.1in}
\parab{Summary:} We believe existing proxy-based solutions need only minor changes to benefit from our results. We focused our work on minimizing the proxy-server latency, and making the client-proxy interaction as efficient as possible with today's cutting-edge networking techniques, and open to the easy adoption of other novel methods as they become available. For the substantial software challenges of making proxies work transparently and securely, and the impact of such proxies on the traffic seen by Web servers, we defer to past work like Flywheel and Free Basics, which has tackled many such problems. 

\section{Related work}
\label{sec:related}


\parab{Measurement work:} Prior work~\cite{he2013next,labovitzAmazon,whoWasAkella} evaluated the use of cloud services like Amazon and Azure by popular Web sites using IP matching, finding that $4\%$ of Alexa's top $1$ million domains use Amazon EC2 or Microsoft Azure. Our results quantify latency to popular services directly \emph{from} these platforms, showing that even services not hosted directly within Amazon or Microsoft data centers are hosted (or at least replicated) nearby.

\parab{Protocols and server-side enhancements:} Our measurement work reveals a vector for deploying a variety of enhancements, as we demonstrate by using BBR~\cite{bbr}. We also leverage known techniques such as larger TCP window sizes~\cite{nanditaWindowSize} and persistent connections~\cite{TCPfastopen}.


Other work~\cite{shandian, vroom, polaris} has observed that Web pages involve complex compute and network dependencies, and can be sped up by re-ordering edges in this dependency graph and/or optimizing bottlenecks. However, this work depends on adoption by Web site operators.

\parab{Cloud-assisted browsers:}
Using proxies to speed up Web browsing is not new. Opera's Mini/Turbo~\cite{operamini,operaturbo} and Google's Flywheel~\cite{agababov2015flywheel} are compression proxies. Amazon Silk~\cite{amazonsilk} offloads processing from thin clients to well-provisioned servers in the cloud. These designs preserve resources at clients, but their impact on latency is not consistently positive, as Sivakumar \textit{et al.}~\cite{sivakumar2014cloud} show in their analysis of a popular cloud-assisted browser. We compare our approach to Flywheel in \S\ref{subsec:simulation}.

WebPro~\cite{sehati2016network}, PARCEL~\cite{parcelSivakumar}, and Cumulus~\cite{netravali2015mahimahi} invoke proxies to batch data fetched from Web servers before delivering it to the clients. However, a single proxy as envisioned in these systems (``a well-provisioned cloud server'' in Cumulus~\cite{netravali2015mahimahi}; a proxy ``implemented on a powerful server'' in PARCEL~\cite{parcelSivakumar}) does not expose the full potential of a proxy-based approach. Unlike other past work, Shandian~\cite{shandian} does indeed suggest that their proxies be placed near Web servers. However, their suggested approach (\ie the Web service operators co-locate Shandian with their reverse-proxies) requires action from Web service operators. Complementary to this work, our focus is on showing that: (a) a small number of cloud proxies suffice to achieve proximity to most Web services, and (b) these can be easily exploited to build an immediately deployable system, without the cooperation of Web service operators.

Lastly, unlike any past work, we also evaluate the competitiveness of \sysname against CDNs, showing that it can achieve performance close to an idealized CDNs.

\parab{Workshop paper:} In our own preliminary work~\cite{bhattacherjee2017cloud}, we presented a small measurement on server consolidation, and a brief result on potential PLT reduction. We extend that work in several substantial ways here: (a) with extensive measurements covering $10\times$ more servers, visualization, and new analysis on how latencies could be further reduced; (b) a design and implementation evaluated across a variety of network conditions and metrics; (c) adding results on transport and compression; (d) modeling CDN- and \sysname-based delivery and comparing to an idealized CDN; and (e) a tighter cost analysis. 
\section{Conclusion}
We present measurements showing the consolidation of Web services in or near a small number of cloud data centers, with only $12$ points of presence in these data centers needed to achieve a median latency of under $13$~ms across the top $1$ million most popular Web pages. We also discuss the potential of a small number of additional sites to lower this collective latency even further.

Based on these measurements, we describe and evaluate the design of a simple proxy-based approach, \sysname, for speeding up the Web today, as well as enabling faster evolution of the Web in the future. We show that \sysname can improve Web performance by more than $50\%$ for users with large latencies to Web servers. Using our experimental data, we also build a simple model for the dependence of Web performance on client-server and last-mile latencies, and show that \sysname can achieve results only modestly worse than a well-optimized content delivery network serving the same pages. \sysname flips the CDN-based model of Web page delivery upside down, establishing user presence near the Web servers, which is substantially easier to achieve and manage than the other way around.

Nevertheless, there is still substantial work required to address trust, security, and interplay with service providers, and explore opportunities to further optimize \sysname. To solicit help in this endeavor, our code and measurement data will be released with the paper.

\section{Acknowledgments}

We thank Balakrishnan Chandrasekaran for his help with geolocation. We also appreciate the generous support from Amazon EC2 and Microsoft Azure in the form of their cloud credits programs -- ``AWS Cloud Credits for Research'' and ``Microsoft Azure Research Award'' respectively, that enabled us to conduct large-scale measurements and experiments.

{\bibliographystyle{abbrv}
\footnotesize
{
\bibliography{cgn_paper}
}
}
\end{document}